\documentclass[
prd,preprint,aps,endfloats,tightenlines,
showpacs,superscriptaddress,nofootinbib,
amssymb
]{revtex4}
\usepackage{graphicx}
\begin{document}
\setlength{\baselineskip}{16pt}
%
%---------------------------------------------------------------------
%
\title{
$\rho$ Meson Decay
in $2+1$ Flavor Lattice QCD
}
%
%---------------------------------------------------------------------------
%
\author{ S.~Aoki }
\affiliation{
Graduate School of Pure and Applied Sciences,
University of Tsukuba,
Tsukuba, Ibaraki 305-8571, Japan
}
\affiliation{
Center for Computational Sciences,
University of Tsukuba,
Tsukuba, Ibaraki 305-8577, Japan
}
%--------------------------------------------
%
\author{ K-I.~Ishikawa }
\affiliation{
Department of Physics,
Hiroshima University,
Higashi-Hiroshima, Hiroshima 739-8526, Japan
}
%--------------------------------------------
%
\author{ N.~Ishizuka }
\affiliation{
Graduate School of Pure and Applied Sciences,
University of Tsukuba,
Tsukuba, Ibaraki 305-8571, Japan
}
\affiliation{
Center for Computational Sciences,
University of Tsukuba,
Tsukuba, Ibaraki 305-8577, Japan
}
%--------------------------------------------
%
\author{ K.~Kanaya }
\affiliation{
Graduate School of Pure and Applied Sciences,
University of Tsukuba,
Tsukuba, Ibaraki 305-8571, Japan
}
%--------------------------------------------
%
\author{ Y.~Kuramashi }
\affiliation{
Graduate School of Pure and Applied Sciences,
University of Tsukuba,
Tsukuba, Ibaraki 305-8571, Japan
}
\affiliation{
Center for Computational Sciences,
University of Tsukuba,
Tsukuba, Ibaraki 305-8577, Japan
}
\affiliation{
RIKEN Advanced Institute for Computational Science,
Kobe, Hyogo 650-0047, Japan
}
%--------------------------------------------
%
\author{ Y.~Namekawa }
\affiliation{
Center for Computational Sciences,
University of Tsukuba,
Tsukuba, Ibaraki 305-8577, Japan
}
%--------------------------------------------
%
\author{ M.~Okawa }
\affiliation{
Department of Physics,
Hiroshima University,
Higashi-Hiroshima, Hiroshima 739-8526, Japan
}
%--------------------------------------------
%
\author{ Y.~Taniguchi }
\affiliation{
Graduate School of Pure and Applied Sciences,
University of Tsukuba,
Tsukuba, Ibaraki 305-8571, Japan
}
\affiliation{
Center for Computational Sciences,
University of Tsukuba,
Tsukuba, Ibaraki 305-8577, Japan
}
%--------------------------------------------
%
\author{ A.~Ukawa }
\affiliation{
Graduate School of Pure and Applied Sciences,
University of Tsukuba,
Tsukuba, Ibaraki 305-8571, Japan
}
\affiliation{
Center for Computational Sciences,
University of Tsukuba,
Tsukuba, Ibaraki 305-8577, Japan
}
%--------------------------------------------
%
\author{ N.~Ukita }
\affiliation{
Center for Computational Sciences,
University of Tsukuba,
Tsukuba, Ibaraki 305-8577, Japan
}
%--------------------------------------------
%
\author{ T.~Yamazaki }
%-- \affiliation{
%-- Center for Computational Sciences,
%-- University of Tsukuba,
%-- Tsukuba, Ibaraki 305-8577, Japan
%-- }
%
\affiliation{
Kobayashi-Maskawa Institute
for the Origin of Particles and the Universe,
Nagoya University, Nagoya, Aichi 464-8602, Japan
}
%--------------------------------------------
%
\author{ T.~Yoshi\'{e} }
\affiliation{
Graduate School of Pure and Applied Sciences,
University of Tsukuba,
Tsukuba, Ibaraki 305-8571, Japan
}
\affiliation{
Center for Computational Sciences,
University of Tsukuba,
Tsukuba, Ibaraki 305-8577, Japan
}
%--------------------------------------------
%
\collaboration{ PACS-CS Collaboration }
%
%---------------------------------------------------------------------------
%
\date{ \today }
%%-- \date{ {\bf RhoD\_PACSCS\_v00 : 110405 : } \today }
%%-- \date{ {\bf RhoD\_PACSCS\_v01 : 110614 : } \today }
%%-- \date{ {\bf RhoD\_PACSCS\_v02 : 110618 : } \today }
%%-- \date{ {\bf RhoD\_PACSCS\_v03 : 110619 : } \today }
%%-- \date{ {\bf RhoD\_PACSCS\_v04 : 110624 : } \today }
%%-- \date{ {\bf RhoD\_PACSCS\_v05 : 110624 : } \today }
%%-- \date{ {\bf RhoD\_PACSCS\_v08 : 110810 : } \today }
%%-- \date{ {\bf RhoD\_PACSCS\_v09 : 111017 : } \today }
%
%---------------------------------------------------------------------------
%
\begin{abstract}
We perform a lattice QCD study of the $\rho$ meson decay
from the $N_f=2+1$ full QCD configurations
generated with a renormalization group improved gauge action
and a non-perturbatively $O(a)$-improved Wilson fermion action.
The resonance parameters,
the effective $\rho\to\pi\pi$ coupling constant
and the resonance mass,
are estimated from the $P$-wave scattering phase shift
for the isospin $I=1$ two-pion system.
The finite size formulas are employed to calculate the phase shift
from the energy on the lattice.
Our calculations are carried out
at two quark masses,
    $m_\pi=410\,{\rm MeV}$ ($m_\pi/m_\rho=0.46$)
and $m_\pi=300\,{\rm MeV}$ ($m_\pi/m_\rho=0.35$),
on a $32^3\times 64$ ($La=2.9\,{\rm fm}$) lattice
at the lattice spacing $a=0.091\,{\rm fm}$.
We compare our results at these two quark masses
with those given in the previous works using
$N_f=2$ full QCD configurations
and the experiment.
\end{abstract}
\maketitle
%
%======================================================================
% @ Introduction
%
\section{ Introduction }
\label{Sec:Introduction}
Recent progress of simulation algorithms,
supported by the development of computer power,
has made it possible to study hadron physics at the physical quark mass
by lattice QCD
(see Ref.~\cite{Rev:DSMI} for recent reviews),
and lattice calculations have clarified the properties of many hadrons.
The studies are mostly concentrated on stable hadrons, however.
Resonances pose an important issue both in terms of
methodologies and physical results.

Among the resonances,
the $\rho$ meson is an ideal case for the lattice calculations,
because the final state of the decay is the two-pion state
which can be treated on the lattice precisely.
In the early stage of studies of the $\rho$ meson decay,
the transition amplitude $\langle \pi\pi|\rho\rangle$
extracted from the time behavior of
the correlation function
$\langle \pi(t)\pi(t) \rho(0) \rangle$
was used to estimate the decay width,
assuming that the hadron interaction
is small enough so that
$\langle \pi\pi|\rho\rangle \ll
| \langle \rho  |\rho   \rangle
  \langle \pi\pi|\pi\pi \rangle |^{1/2}$
is satisfied~\cite{rhd:TAMP:GMTW,rhd:TAMP:LD,rhd:TAMP:MM,rhd:TAMP:JMMU}.

A more realistic approach is a study
from the $P$-wave scattering phase shift
for the isospin $I=1$ two-pion system.
The finite size formulas
presented by L\"uscher in the center of mass frame~\cite{Lfm:L}
and extensions to non-zero total momentum frames~\cite{Lfm:RG,Lfm:ETMC}
are employed for an estimation of the phase shift
from an eigenvalue of the energy on the lattice.
The first study of this approach
was carried out by CP-PACS Collaboration
using $N_f=2$ full QCD configurations
($m_\pi=330\,{\rm MeV}$, $a=0.21\,{\rm fm}$,
$La=2.5\,{\rm fm}$)~\cite{rhd:SCPH:CP-PACS}.
After this work
ETMC Collaboration presented results
with $N_f=2$ configurations
at several quark masses
($m_\pi=290,330\,{\rm MeV}$ ($La=2.5\,{\rm fm}$),
 $m_\pi=420,480\,{\rm MeV}$ ($La=1.9\,{\rm fm}$),
$a=0.079\,{\rm fm}$)~\cite{rhd:SCPH:ETMC_1,rhd:SCPH:ETMC_2}.
Recently Lang {\it et al.} reported
results of high statistical calculations
with the Laplacian Heaviside smearing operators
on a single $N_f=2$ gauge ensemble
($m_\pi=266\,{\rm MeV}$, $a=0.124\,{\rm fm}$,
$La=1.98\,{\rm fm}$)~\cite{rhd:SCPH:LANG}.

In the present work
we extend these studies by employing
$N_f=2+1$ full QCD configurations
and working on a larger lattice volume.
Our calculations are carried out with
the gauge configurations previously generated
by PACS-CS Collaboration
with a renormalization group improved gauge action
and a non-perturbatively $O(a)$-improved Wilson fermion action
at $\beta=1.9$ on $32^3\times 64$ lattice
($a=0.091\,{\rm fm}$, $La=2.9\,{\rm fm}$)~\cite{conf:PACS-CS}.
We choose two subsets of the PACS-CS configurations.
One of them corresponds to the hopping parameters
$\kappa_{ud}=0.13754$ for the degenerate up and down quarks and
$\kappa_{s }=0.13640$ for the strange quark,
for which the pion mass takes
$m_\pi=410\,{\rm MeV}$ ($m_\pi/m_\rho=0.46$).
The other is at
$\kappa_{ud}=0.13770$ and $\kappa_{s }=0.13640$,
corresponding to $m_\pi=300\,{\rm MeV}$ ($m_\pi/m_\rho=0.35$).
We extract the scattering phase shift
on three momentum frames,
the center of mass and
the non-zero momentum frames
with the total momentum ${\bf P}=(2\pi/L)(0,0,1)$
and ${\bf P}=(2\pi/L)(1,1,0)$,
to obtain the phase shifts
at various energies from a single full QCD ensemble
as in the previous works by
ETMC~\cite{rhd:SCPH:ETMC_1,rhd:SCPH:ETMC_2}
and Lang {\it et al.}~\cite{rhd:SCPH:LANG}.

We note that
QCDSF Collaboration calculated
the scattering phase shifts for the ground state
in the center of mass frame at several quark masses.
($m_\pi=240-430\,{\rm MeV}$)~\cite{rhd:SCPH:QCDSF}.
They estimated the resonance parameters from these results,
assuming that the effective $\rho\to\pi\pi$ coupling constant
does not depend on the quark mass.
BMW Collaboration
presented their first preliminary results
with $N_f=2+1$ configurations
($m_\pi=200, 340\,{\rm MeV}$, $a=0.116\,{\rm fm}$)
at Lattice 2010~\cite{rhd:SCPH:BMW}.
We also refer to works exploring
an application of the stochastic Laplacian Heaviside smearing
to the two-pion states with the isospin $I=0,1,2$
using $N_f=2+1$ configurations on the large lattice volume
in Ref.~\cite{rhd:try:HSC}.

This paper is organized as follows.
In Sec.~\ref{Sec:Methods}
we give the method of the calculation.
The simulation parameters of the present work are also presented.
We present our results
and compare ours with those by the other works
in Sec.~\ref{Sec:Results}.
Conclusions of the present work are given in Sec.~\ref{Sec:Conclusions}.
Result of a pilot study of the present work
at $m_\pi=410\,{\rm MeV}$
was reported at Lattice 2010~\cite{rhd:SCPH:PACS-CS}.
All calculations are carried out
on the PACS-CS computer at Center for Computational Sciences,
University of Tsukuba.
%
%======================================================================
%
\section{ Methods }
\label{Sec:Methods}
%
%----------------------------------------------------------------------
% @ Finite size formula
%
\subsection{ Finite size formula }
In order to calculate
the $P$-wave scattering phase shift
for the isospin $I=1$ two-pion system at various energies
from a single full QCD ensemble,
we consider three momentum frames,
the center of mass frame (CMF),
the non-zero momentum frames
with total momentum ${\bf P}=(2\pi/L)(0,0,1)$ (MF1)
and ${\bf P}=(2\pi/L)(1,1,0)$ (MF2).
These frames have been also considered in the previous works by
ETMC~\cite{rhd:SCPH:ETMC_1,rhd:SCPH:ETMC_2}
and Lang {\it et al.}~\cite{rhd:SCPH:LANG}.
In these momentum frames the $P$-wave state is decomposed as
\begin{equation}
\begin{array}{lc cc lc lc }
\mbox{frame} &\ & {\bf P} L/(2\pi) &\ & {\rm g} &\ & \Gamma              \\
\mbox{CMF}   &  & (0,0,0)  && O_h     && {\bf T}_{1}^{-}                 \\
\mbox{MF1}   &  & (0,0,1)  && D_{4h}  && {\bf A}_{2}^{-} + {\bf E}^{-}   \\
\mbox{MF2}   &  & (1,1,0)  && D_{2h}  && {\bf B}_{1}^{-}
                                       + {\bf B}_{2}^{-}
                                       + {\bf B}_{3}^{-}    \\
\end{array}
\ \ ,
\label{eq:P-wave_decomp}
\end{equation}
where ${\bf P}$ is the total momentum,
${\rm g}$ is the rotational group in each momentum frame on the lattice
and $\Gamma$ is the irreducible representation of the rotational group.
In the present work
we consider four irreducible representations :
${\bf T}_{1}^{-}$ in the CMF,
${\bf A}_{2}^{-}$ and ${\bf E}^{-}$ in the MF1, and
${\bf B}_{1}^{-}$ in the MF2.
The ground and the first excited states of these representations
with the isospin $(I,I_z)=(1,0)$,
ignoring the hadron interactions,
are shown in Table~\ref{table:calc_rep}.

The scattering phase shift is related to
an eigenvalue of the energy on the lattice by the finite size formula.
The formula in the CMF was presented by L\"uscher~\cite{Lfm:L},
that in the MF1 by Rummukainen and Gottlieb~\cite{Lfm:RG},
and the MF2 by ETMC~\cite{Lfm:ETMC}.
The formulas for the representations considered in the present work
are written by
\begin{equation}
\frac{1}{\tan\delta(k)} =
\left\{
\begin{array}{l l }
V_{00}                & \quad \mbox{ for \,${\bf T}_{1}^{-}$\, in CMF }  \\
V_{00} -     V_{20}   & \quad \mbox{ for \,${\bf E}^{-}    $\, in MF1 }  \\
V_{00} + 2\, V_{20}   & \quad \mbox{ for \,${\bf A}_{2}^{-}$\, in MF1 }  \\
V_{00} -     V_{20} + \sqrt{6} \, V_{22}
                      & \quad \mbox{ for \,${\bf B}_{1}^{-}$\, in MF2 }
\end{array}
\right.
\ ,
\label{eq:FSF_our_rep}
\end{equation}
with the $P$-wave scattering phase shift $\delta(k)$.
The real function $V_{lm}$
is defined by
\begin{equation}
  V_{lm}(k;{\bf P})
     = \frac{1}{\gamma q^{l+1}}
       \frac{1}{\pi^{3/2} \sqrt{ 2l +1 } }
       \, e^{ - i m \pi /4 }
       \cdot Z_{lm}( 1 ; q ; {\bf d} )
\ ,
\label{eq:def_Vlm}
\end{equation}
with ${\bf d}={\bf P}L/(2\pi)$ and $q=kL/(2\pi)$,
where
${\bf P}$ is the total momentum and
$k$ is the two-pion scattering momentum
defined from the invariant mass
$\sqrt{s}$ by
$\sqrt{s} = \sqrt{ E^2 - |{\bf P}|^2 } = 2\sqrt{k^2 + m_\pi^2}$
with the energy $E$ in the non-zero momentum frame.
In (\ref{eq:def_Vlm})
$\gamma$ is the Lorentz boost factor from the non-zero momentum frame
to the center of mass frame given by $\gamma = E / \sqrt{s}$.
The function $Z_{lm}(1;q;{\bf d})$ in (\ref{eq:def_Vlm})
is an analytic continuation of
\begin{equation}
   Z_{lm}( s ; q ; {\bf d})
   = \sum_{{\bf r}\in{D({\bf d})}}
     \, {\cal Y}_{lm}({\bf r}) \cdot ( |{\bf r}|^2 - q^2 )^{-s}
\ ,
\label{eq:def_Zlm}
\end{equation}
which is defined for ${\rm Re}(s) > (l+3)/2$, where
${\cal Y}_{lm}({\bf r})$ is a polynomial
related to the spherical harmonics
through ${\cal Y}_{lm}({\bf r}) = |{\bf r}|^l \cdot Y_{lm}(\Omega)$
with $\Omega$ the spherical coordinate for ${\bf r}$.
The convention of $Y_{lm}(\Omega)$ is that of~\cite{book:MESSIAH}.
The summation for ${\bf r}$ in (\ref{eq:def_Zlm}) runs over the set,
\begin{equation}
  D({\bf d})
    = \left\{
       \,  {\bf r}
       \,  |
       \,  {\bf r} = \hat{\gamma}^{-1}
                        ( {\bf n} + {\bf d}/2 )  \, ,
       \,\,  {\bf n} \in \mathbb{Z}^3
       \, \right\}
\ .
\end{equation}
The operation $\hat{ \gamma }^{ -1 }$ is the inverse Lorentz transformation :
$\hat{\gamma}^{-1} {\bf x} = {\bf x}_{||} /\gamma + {\bf x}_{\bot}$ ,
where $ {\bf x}_{||} = ( {\bf x} \cdot {\bf d} ) {\bf d} / {\bf d}^2$
is the parallel component and
${\bf x}_{\bot} = {\bf x} - {\bf x}_{||}$ is
the perpendicular component of the vector ${\bf x}$ in the direction ${\bf d}$.
$Z_{lm}(1;q;{\bf d})$ can be evaluated
by the method described in Ref.~\cite{I2_phsh:Yamazaki}.
%
%-------------------------------------------------------------------------------
% @ Extraction of energies
%
\subsection{ Extraction of energies }
\begin{figure}[h]
\includegraphics[width=11.0cm]{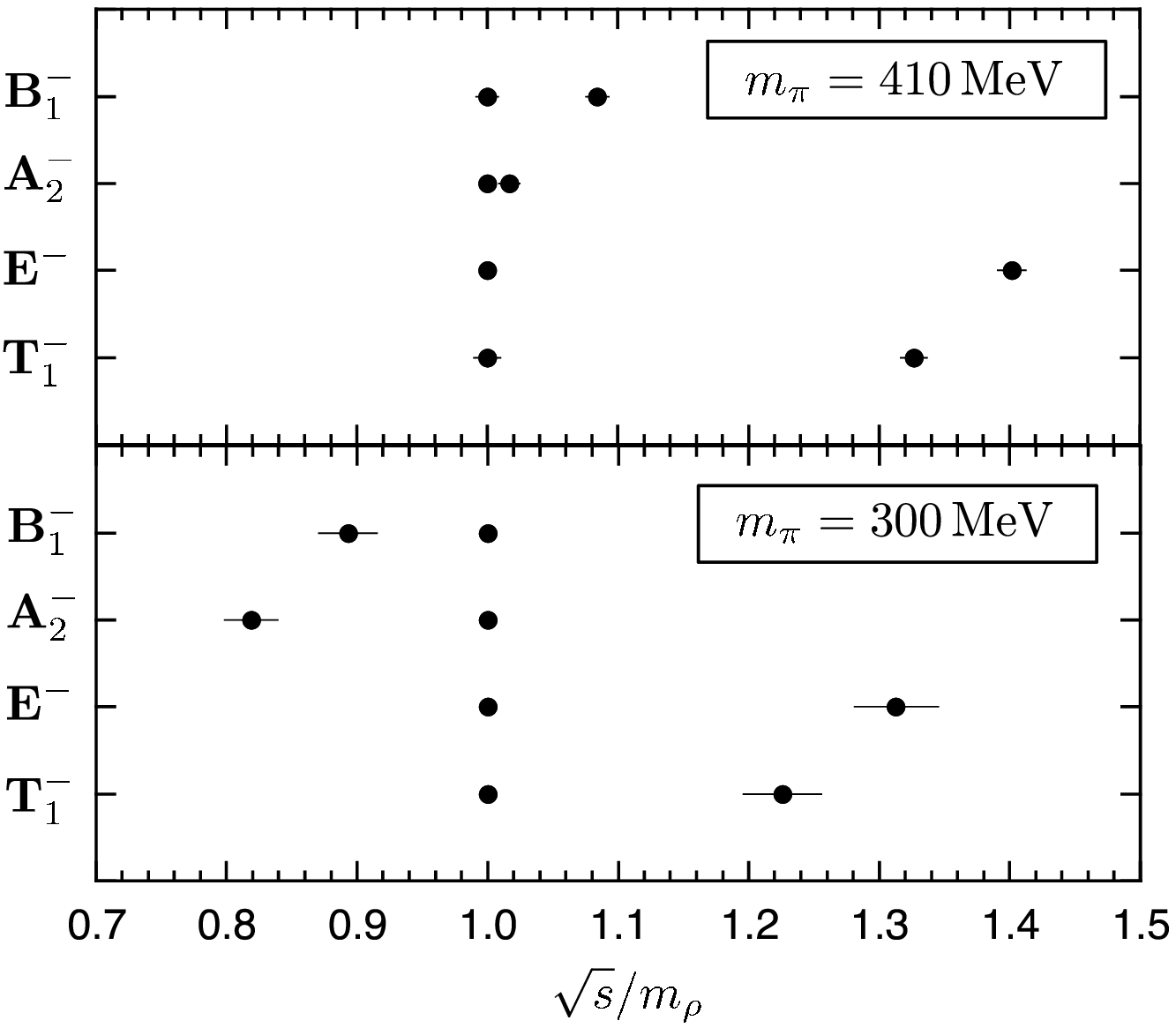}
\caption{
$\sqrt{s}/m_\rho$
for states tabulated in Table~\ref{table:calc_rep}
on our gauge configurations at
$m_\pi=410\,{\rm MeV}$ (upper panel)
and $m_\pi=300\,{\rm MeV}$ (lower panel).
}
\label{fig:kinematics}
\end{figure}

In Fig.~\ref{fig:kinematics}
we show values of the invariant mass $\sqrt{s}$
divided by the $\rho$ meson mass $m_\rho$ for
the states tabulated in Table~\ref{table:calc_rep}
on our gauge configurations at
$m_\pi=410\,{\rm MeV}$ (upper panel) and
$m_\pi=300\,{\rm MeV}$ (lower panel).
Here
we ignore the hadron interactions and use
the values of $m_\rho$ and $m_\pi$
obtained in the previous work in Ref.~\cite{conf:PACS-CS}.
For the ${\bf T}^{-}_1$ and the ${\bf E}^{-}$ representation,
we only calculate the scattering phase for the ground state
in the present work.
From Fig.~\ref{fig:kinematics}
the energies of these states
are expected to be much smaller
than those of the excited states,
even if the hadron interaction is switched on.
Thus, we extract the energies of these states
by a single exponential fit for
the time correlation functions of the $\rho$ meson
as carried out in a usual study of the hadron spectrum.
We use the local $\rho$ meson operator for the sink and
a smeared operator
for the source
as discussed later.

For the ${\bf A}_{2}^{-}$ and the ${\bf B}_{1}^{-}$ representation,
we also calculate the scattering phase shift
for the first excited state.
In order to extract the energies of
the lower two states for these representations,
we use the variational method~\cite{method_diag:LW}
with a matrix of the time correlation function,
\begin{equation}
G(t) = \left(
\begin{array}{ll}
       \langle 0|\, (\pi\pi)^\dagger(t)\, \overline{(\pi\pi)}(t_s)\, |0\rangle
& \,\, \langle 0|\, (\pi\pi)^\dagger(t)\, \overline{\rho    }(t_s)\, |0\rangle  \\
       \langle 0|\,     \rho^\dagger(t)\, \overline{(\pi\pi)}(t_s)\, |0\rangle
& \,\, \langle 0|\,     \rho^\dagger(t)\, \overline{\rho    }(t_s)\, |0\rangle
\end{array}
\right)
\ ,
\label{eq:def_G}
\end{equation}
for each representation.
The energies are extracted
from two eigenvalues $\lambda_n (t)$ ($n=1,2$) of the matrix,
\begin{equation}
  M(t) = G(t)\,  G^{-1}(t_R)
\ ,
\label{eq:def_M}
\end{equation}
with some reference time $t_R$.

Here we comment on the discussion
of the generalized eigenvalue problem (GEVP)
by ALPHA Collaboration in Ref.~\cite{method_GEVP}.
In a $2\times 2$ matrix case of $G(t)$,
they proved that the effective mass
of the eigenvalue $\lambda_n (t)$ ($n=1,2$)
of the matrix $M(t)$ in (\ref{eq:def_M})
can be written as
\begin{equation}
  E_n + O( {\rm e}^{ - (E_m - E_n) t } )
\qquad ( n=1,2 \ , \ m \ge 3 )
\ ,
\label{eq:GEVP_col}
\end{equation}
in a large time region
with the eigenvalue of the energy $E_i$ ($i=1,2,\cdots$ ).
Here it should be noted that
their proof was given only for the case
where $G(t)$ is a Hermitian matrix.
In our case we use different operators
for the sink and source in $G(t)$
as explained later.
Therefore $G(t)$ is not a Hermitian matrix and
the discussion of GEVP cannot be applied to our case.
In the present work
we assume that the lower two states dominate $G(t)$
in a large time region.
This is expected to be a good approximation,
because
the second excited state takes a much higher value
( $E_3/E_2>1.3$ in the absence of hadron interactions
for both quark masses studied
in the present work ).
With this assumption,
the second term of (\ref{eq:GEVP_col}) does not appear
for a general case of $G(t)$,
and the energy for the ground and the first excited states
can be extracted by a single exponential fit
for the eigenvalue $\lambda_n (t)$ ($n=1,2$)
in a large $t$ region.

In (\ref{eq:def_G})
the operator $\rho(t)$ is given by
\begin{eqnarray}
  \rho(t) = \sum_{j=1}^{3}\, p_j \cdot \rho_j({\bf p},t) / |{\bf p}|
\ ,
\end{eqnarray}
where $\rho_j({\bf p},t)$ is
the local operator
for the neutral $\rho$ meson at the time slice $t$ with the momentum ${\bf p}$.
The momentum takes
${\bf p}=(2\pi/L)(0,0,1)$ for the ${\bf A}_{2}^{-}$ and
${\bf p}=(2\pi/L)(1,1,0)$ for the ${\bf B}_{1}^{-}$ representation.
Hereafter
we assume that the momentum ${\bf p}$
takes one of these two values depending on the representation.

In (\ref{eq:def_G}) $(\pi\pi)(t)$  is an operator
for the two pions with the momentum ${\bf 0}$ and ${\bf p}$,
which is defined by
\begin{equation}
  (\pi\pi)(t) = \frac{1}{\sqrt{2}}
    \left(    \pi^{+}({\bf 0},t_1) \, \pi^{-}({\bf p},t)
            - \pi^{-}({\bf 0},t_1) \, \pi^{+}({\bf p},t)
    \right) \times {\rm e}^{ m_\pi \cdot ( t_1 - t ) }
\ ,
\label{eq:pp_op_sink}
\end{equation}
where
$\pi^{\pm}({\bf p},t)$ is the local pion operator
with the momentum ${\bf p}$ at the time slice $t$.
The time slice of the pion with the zero momentum
is fixed at $t_1 \gg t$,
and the time slice of the other pion $t$ runs over the whole time extent.
An exponential time factor
in (\ref{eq:pp_op_sink}) is introduced
so that the operator
has the same time behavior as that of the usual Heisenberg operator,
{\it i.e.},
\begin{equation}
    \langle 0 | \, (\pi\pi)^\dagger (t)
  = \langle 0 | \, (\pi\pi)^\dagger (0) \, {\rm e}^{ - H t }
\qquad \mbox{ for $t_1 \gg t$ }  \ ,
\end{equation}
with the Hamiltonian $H$.

In the previous works
the time slices of the two operators for the pion
in the sink operator (\ref{eq:pp_op_sink})
are set equal $t_1=t$,
and they simultaneously
run over some time interval.
In that case
we need to repeat solving quark propagators for the time slices
in that time interval.
This computer-time consuming procedure can be avoided by fixing
the time slice of one of the pion at $t_1$ as done in the present work.
But we need to set $t_1 \gg t$ to avoid contamination
from higher energy states produced by the operator at $t_1$
in this method.

Two operators
$\overline{(\pi\pi)}(t_s)$ and
$\overline{\rho}(t_s)$
are used for the sources in (\ref{eq:def_G}),
which are given by
\begin{eqnarray}
&&
\overline{(\pi\pi)}(t_s)
= \frac{1}{\sqrt{2}}
   \Bigl(   \pi^{+}({\bf 0},t_s) \pi^{-}({\bf p},t_s)
          - \pi^{-}({\bf 0},t_s) \pi^{+}({\bf p},t_s)   \Bigl)
\ ,
\label{eq:pp_op_source}
\\
&&
\overline{\rho}(t_s) =
  \frac{1}{N_\Gamma}
  \sum_{{\bf z}\in \Gamma}
     \frac{1}{\sqrt{2}}
      \Bigl(   \overline{U}({\bf z},t_s) \gamma_p U({\bf z},t_s)
             - \overline{D}({\bf z},t_s) \gamma_p D({\bf z},t_s)   \Bigl)
   {\rm e}^{ i {\bf p}\cdot{\bf z} }
\ ,
\label{eq:rho_op_source}
\end{eqnarray}
where $\gamma_p=\sum_{j=1}^{3} p_j \cdot \gamma_j /|{\bf p}|$.
The operator
$Q({\bf z},t_s)$ ($Q=U,D$)
is a smeared operator
for the up or the down quark given by
\begin{equation}
  Q({\bf z},t_s) = \sum_{\bf_x} q({\bf x},t_s )
                   \cdot \Psi(|{\bf x} - {\bf z}|)
\ ,
\label{eq:smearing_quark}
\end{equation}
where $q({\bf x},t_s)$ ($q=u,d)$ is the up or the down quark operator
at the position ${\bf x}$ and the time $t_s$.
We adopt the same smearing function
$\Psi(|{\bf x}|)$ as in Ref.~\cite{conf:PACS-CS},
{\it i.e.},
$\Psi(|{\bf x}|) = A \exp( - B |{\bf x}| )$
with $\Psi(0)=1$ and the smearing parameters :
$(A,B)=(1.2,0.17)$ for $m_\pi=410\,{\rm MeV}$ and
$(A,B)=(1.2,0.09)$ for $m_\pi=300\,{\rm MeV}$.
The operator (\ref{eq:smearing_quark})
is used after fixing gauge configurations
to the Coulomb gauge,
assuming that an ambiguity of the gauge fixing
does not cause a significant systematic error
in the time correlation function.
In (\ref{eq:rho_op_source})
a summation over ${\bf z}$ is taken
to reduce a statistical error
and
\begin{equation}
  \Gamma = \{\ {\bf z}\ |\ {\bf z}=(L/2)\cdot(n_1,n_2,n_3)
  \ , \ n_j = \mbox{$0$ or $1$}\ , \ N_\Gamma = 8 \ \}
\label{eq:def_smearing_Gamma}
\end{equation}
is chosen in the present work.
The smeared operator
(\ref{eq:rho_op_source})
is also used to extract the energy of the ground state for
the ${\bf T}_1^-$ and
the ${\bf E}^{-}$ representation,
setting the momentum ${\bf p}={\bf 0}$ and ${\bf p}=(2\pi/L)(0,0,1)$,
respectively.

Here we note that
the operator
$(\pi\pi)(t)$ in (\ref{eq:pp_op_sink})
is not an eigenstate under exchange of the momenta of the two pions.
Thus it has no definite parity and
does not belong to the irreducible representation of
the $D_{4h}$ for ${\bf p}=(2\pi/L)(0,0,1)$ and
the $D_{2h}$ for ${\bf p}=(2\pi/L)(1,1,0)$.
It is actually a linear combination of
components of two irreducible representations
${\bf A}_2^{-} + {\bf A}_1^{+}$ for the $D_{4h}$ and
${\bf B}_1^{-} + {\bf A}_1^{+}$ for the $D_{2h}$.
The other operators in $G(t)$ in (\ref{eq:def_G}), however,
belong to the irreducible representation
${\bf A}_2^{-}$ or ${\bf B}_1^{-}$ depending on the momentum ${\bf p}$.
Therefore, it is not necessary to worry about
mixing to other irreducible representations
in $G(t)$ in (\ref{eq:def_G}).
%
%-------------------------------------------------------------------------------
% @ calc of G(t)
%
\subsection{ Calculation of $G(t)$ }
\begin{figure}[h]
\includegraphics[width=14.5cm]{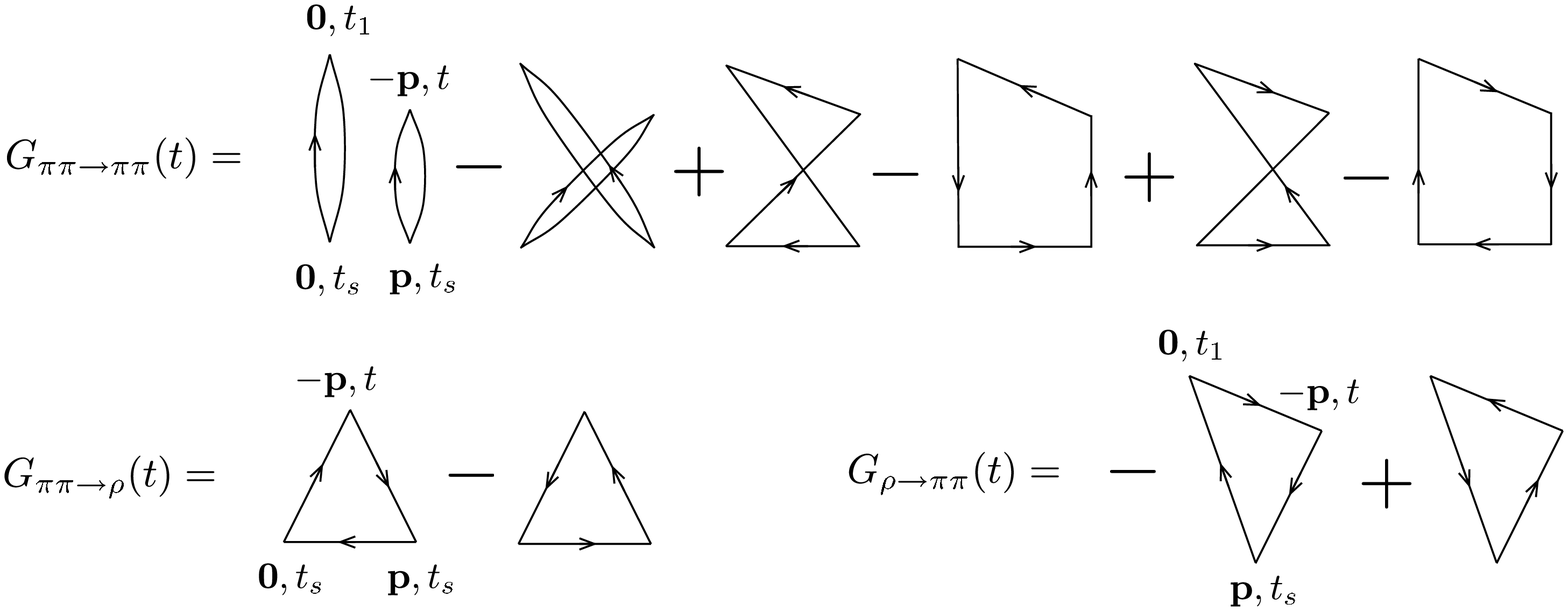}
\caption{
Quark contractions of
$\pi\pi\to\pi\pi$, $\pi\pi\to\rho$ and
$\rho\to\pi\pi$ components of
the matrix of the time correlation function $G(t)$.
The time runs upward in the diagrams.
The vertices refer to the pion or the $\rho$ meson operator
with the momentum at the time slice specified in the diagrams.
The momentum ${\bf p}$
takes ${\bf p}=(2\pi/L)(0,0,1)$ for the ${\bf A}_{2}^{-}$
and   ${\bf p}=(2\pi/L)(1,1,0)$ for the ${\bf B}_{1}^{-}$
representation.
}
\label{fig:quark_cont_G}
%------------
\newpage
%------------
\end{figure}

The quark contractions of the components of
the matrix of the time correlation function
$G(t)$ in (\ref{eq:def_G})
are shown
in Fig.~\ref{fig:quark_cont_G}.
The time runs upward in the diagrams.
The vertices refer to the pion or the $\rho$ meson operator
with the momentum at the time slice specified in the diagrams.
The momentum  ${\bf p}$
takes ${\bf p}=(2\pi/L)(0,0,1)$ for the ${\bf A}_{2}^{-}$
  and ${\bf p}=(2\pi/L)(1,1,0)$ for the ${\bf B}_{1}^{-}$ representation.
The $\rho$ meson operators at $t_s$ are
%%@T5-- the smearing operators
the smeared operators
and the other is the local operator.

We calculate the quark contractions in Fig.~\ref{fig:quark_cont_G}
by the source method and the stochastic noise method
as in the previous work by CP-PACS~\cite{rhd:SCPH:CP-PACS}.
We introduce a $U(1)$ noise $\xi_j({\bf x})$
which satisfies
\begin{equation}
 \sum_{j=1}^{N_R}
 \xi_j^\dagger ({\bf x}) \xi_j ({\bf y})
  = \delta^3( {\bf x} - {\bf y} )
             \qquad \mbox{ for \ $N_R \to \infty$ }
\ ,
\label{eq:xi_prop}
\end{equation}
where $N_R$ is the number of noises.
We calculate the following four types of quark propagators:
\begin{eqnarray}
&&
  Q_{AB}( {\bf x}, t | {\bf q}, t_s, \xi_j )
    = \sum_{\bf y} ( D^{-1} )_{AB}({\bf x}, t ; {\bf y}, t_s )
      \cdot \Bigl[ {\rm e}^{ i{\bf q}\cdot{\bf y}} \xi_j({\bf y}) \Bigr]
\ ,
\label{eq:QP_Q}
\\
&&
  W_{AB} ( {\bf x}, t | {\bf k}, t_a | {\bf q}, t_s, \xi_j )
  = \sum_{\bf y}
    \sum_{C}
       ( D^{-1} )_{AC}({\bf x}, t ; {\bf y}, t_a )
         \cdot
         \Bigl[ {\rm e}^{ i{\bf k}\cdot{\bf y}}
                      \gamma_5 \ Q( {\bf y}, t_a | {\bf q}, t_s, \xi_j )
         \Bigr]_{CB}
\ ,
\label{eq:QP_W}
\\
&&
  \overline{Q}_{AB}( {\bf x}, t | {\bf z}, t_s )
    = \sum_{\bf y} ( D^{-1} )_{AB}({\bf x}, t ; {\bf y}, t_s )
      \cdot \Bigl[ \Psi(|{\bf y}-{\bf z}|) \Bigr]
\ ,
\label{eq:QP_Q_SM}
\\
&&
  \overline{W}_{AB} ( {\bf x}, t | {\bf k}, t_a | {\bf z}, t_s )
  = \sum_{\bf y}
    \sum_{C}
       ( D^{-1} )_{AC}({\bf x}, t ; {\bf y}, t_a )
         \cdot
         \Bigl[ {\rm e}^{ i{\bf k}\cdot{\bf y}}
                      \gamma_5 \ \overline{Q}( {\bf y}, t_a | {\bf z}, t_s )
         \Bigr]_{CB}
\ ,
\label{eq:QP_W_SM}
\end{eqnarray}
where $A$, $B$ and $C$ refer to color and spin indices,
and $\Psi(|{\bf x}|)$ is the smearing function in (\ref{eq:smearing_quark}).
The square bracket
is used as the source
for the inversion of the Dirac operator $D$.
The propagators for the smeared quarks
($\overline{Q}$ and $\overline{W}$)
are solved on the gauge configurations
fixed to the Coulomb gauge,
while the gauge is not fixed
for those of the stochastic quarks ($Q$ and $W$).

The function $G_{\pi\pi\to\pi\pi}(t)$
for the first diagram in Fig.~\ref{fig:quark_cont_G}
can be calculated by introducing an another $U(1)$ noise $\eta_j({\bf x})$
having the same property as $\xi_j({\bf x})$ in (\ref{eq:xi_prop}),
\begin{eqnarray}
G^{\rm [1st]}_{\pi\pi\to\pi\pi}(t) &=&
\sum_{j=1}^{N_R}
\sum_{{\bf x},{\bf y}}
\Bigl\langle
     Q^\dagger ( {\bf x}, t_1 | {\bf 0}, t_s, \eta_j  )
  \, Q         ( {\bf x}, t_1 | {\bf 0}, t_s, \eta_j  )
\Bigr\rangle
\, {\rm e}^{ m_\pi \cdot ( t_1 - t ) }
\cr
&&
\quad
\times\,\, {\rm e}^{ - i {\bf p}\cdot{\bf y}}
\Bigl\langle
       Q^\dagger ( {\bf y}, t | {\bf 0}, t_s, \xi_j )
    \, Q         ( {\bf y}, t | {\bf p}, t_s, \xi_j )
\Bigr\rangle
\ ,
\label{eq:D1_quark_c}
\end{eqnarray}
where the bracket means trace for the color and the spin indices.
The exponential time factor comes from the definition
of the operator for the two pions in (\ref{eq:pp_op_sink}).
The $G_{\pi\pi\to\pi\pi}(t)$ for the second diagram
is given by exchanging the momentum and the time slice
of the sink in (\ref{eq:D1_quark_c}).

The function $G_{\pi\pi\to\pi\pi}(t)$ for
the 3rd to 6th diagrams can be obtained by
\begin{eqnarray}
%------------------------
%  B1 = 3 * 5
G^{\rm [3rd]}_{\pi\pi\to\pi\pi}(t) &=&
\sum_{j=1}^{N_R}
\sum_{\bf x} {\rm e}^{ - i {\bf p}\cdot{\bf x}}
\Bigl\langle
      W^\dagger ({\bf x},t| {\bf 0}, t_1 | {\bf 0}, t_s, \xi_j )
  \,  W         ({\bf x},t| {\bf 0}, t_s | {\bf p}, t_s, \xi_j )
\Bigr\rangle
{\rm e}^{ m_\pi \cdot ( t_1 - t ) }
\ ,
\cr
%------------------------
%  B2 = 3 * 4
G^{\rm [4th]}_{\pi\pi\to\pi\pi}(t) &=&
\sum_{j=1}^{N_R}
\sum_{\bf x} {\rm e}^{ - i {\bf p}\cdot{\bf x}}
\Bigl\langle
      W^\dagger ({\bf x},t| {\bf 0}, t_1 | {\bf 0}, t_s, \xi_j )
  \,  W         ({\bf x},t| {\bf p}, t_s | {\bf 0}, t_s, \xi_j )
\Bigr\rangle
{\rm e}^{ m_\pi \cdot ( t_1 - t ) }
\ ,
\cr
%------------------------
%  B3 = 7 * 3
G^{\rm [5th]}_{\pi\pi\to\pi\pi}(t) &=&
\sum_{j=1}^{N_R}
\sum_{\bf x} {\rm e}^{ - i {\bf p}\cdot{\bf x}}
\Bigl\langle
      W         ({\bf x},t| {\bf 0}, t_1 |  {\bf 0}, t_s, \xi_j )
  \,  W^\dagger ({\bf x},t| {\bf 0}, t_s | -{\bf p}, t_s, \xi_j )
\Bigr\rangle
{\rm e}^{ m_\pi \cdot ( t_1 - t ) }
\ ,
\cr
%------------------------
%  B4 = 6 * 3
G^{\rm [6th]}_{\pi\pi\to\pi\pi}(t) &=&
\sum_{j=1}^{N_R}
\sum_{\bf x} {\rm e}^{ - i {\bf p}\cdot{\bf x}}
\Bigl\langle
      W         ({\bf x},t|  {\bf 0}, t_1 |  {\bf 0}, t_s, \xi_j )
  \,  W^\dagger ({\bf x},t| -{\bf p}, t_s |  {\bf 0}, t_s, \xi_j )
\Bigr\rangle
{\rm e}^{ m_\pi \cdot ( t_1 - t ) }
\ .
\label{eq:QC_B}
\end{eqnarray}
The function $G_{\pi\pi\to\rho}(t)$ for two diagrams
in Fig.~\ref{fig:quark_cont_G}
can be similarly calculated by
\begin{eqnarray}
%---------------------
%  Y1 = 6 * 0
G^{\rm [1st]}_{\pi\pi\to\rho}(t) &=&
\sum_{j=1}^{N_R}
\sum_{\bf x}  {\rm e}^{ - i {\bf p}\cdot{\bf x}}
\Bigl\langle
       W^\dagger ({\bf x},t| - {\bf p}, t_s | {\bf 0}, t_s, \xi_j )
   \,  (\gamma_5 \gamma_p )
   \,  Q         ({\bf x},t                 | {\bf 0}, t_s, \xi_j )
\Bigr\rangle
\ ,
\cr
%---------------------
%  Y2 = 0 * 4
G^{\rm [2nd]}_{\pi\pi\to\rho}(t) &=&
\sum_{j=1}^{N_R}
\sum_{\bf x}  {\rm e}^{ - i {\bf p}\cdot{\bf x}}
\Bigl\langle
       Q^\dagger ({\bf x},t                | {\bf 0}, t_s, \xi_j )
  \,   (\gamma_5 \gamma_p )
  \,   W         ({\bf x},t | {\bf p}, t_s | {\bf 0}, t_s, \xi_j )
\Bigr\rangle
\ ,
\label{eq:QC_pprh}
\end{eqnarray}
where
$\gamma_p = \sum_{j=1}^{3} p_j \cdot \gamma_j / |{\bf p}|$.

We can obtain the function $G_{\rho\to\pi\pi}(t)$
for two diagrams in Fig.~\ref{fig:quark_cont_G}
by
\begin{eqnarray}
%-------------------
%  X1 = 0 * 1
G^{\rm [1st]}_{\rho\to\pi\pi}(t) &=&
-
\frac{1}{N_\Gamma}
\sum_{{\bf z}\in\Gamma} {\rm e}^{   i {\bf p}\cdot{\bf z}}
\sum_{\bf x}            {\rm e}^{ - i {\bf p}\cdot{\bf x}}
\Bigl\langle
      \overline{Q}^\dagger ( {\bf x}, t                | {\bf z}, t_s )
  \,  \overline{W}         ( {\bf x}, t | {\bf 0}, t_1 | {\bf z}, t_s )
  \,  (\gamma_5 \gamma_p )
\Bigr\rangle
{\rm e}^{ m_\pi \cdot ( t_1 - t ) }
\ ,
\cr
%-------------------
%  X2 = 1 * 0
G^{\rm [2nd]}_{\rho\to\pi\pi}(t) &=&
-
\frac{1}{N_\Gamma}
\sum_{{\bf z}\in\Gamma} {\rm e}^{   i {\bf p}\cdot{\bf z}}
\sum_{\bf x}            {\rm e}^{ - i {\bf p}\cdot{\bf x}}
\Bigl\langle
      \overline{W}^\dagger ( {\bf x}, t | {\bf 0}, t_1 | {\bf z}, t_s )
  \,  \overline{Q}         ( {\bf x}, t                | {\bf z}, t_s )
  \,  (\gamma_5 \gamma_p )
\Bigr\rangle
{\rm e}^{ m_\pi \cdot ( t_1 - t ) }
\ ,
\label{eq:QC_rhpp}
\end{eqnarray}
where $\Gamma$ and $N_\Gamma$ are defined by (\ref{eq:def_smearing_Gamma}).
The component $G_{\rho\to\rho}(t)$ is given
by the $\overline{Q}$-type propagators
as the usual time correlation function for the $\rho$ meson,
\begin{equation}
G_{\rho\to\rho}(t) =
-
\frac{1}{N_\Gamma}
\sum_{{\bf z}\in\Gamma} {\rm e}^{   i {\bf p}\cdot{\bf z}}
\sum_{\bf x} {\rm e}^{ - i {\bf p}\cdot{\bf x}}
\Bigl\langle
      \overline{Q}^\dagger ( {\bf x}, t | {\bf z}, t_s )
      (\gamma_5 \gamma_p )
  \,  \overline{Q}         ( {\bf x}, t | {\bf z}, t_s )
  \,  (\gamma_5 \gamma_p )
\Bigr\rangle
\ .
\label{eq:QC_rhrh}
\end{equation}

We calculate the $Q$-type propagators (\ref{eq:QP_Q})
for combinations of ${\bf q}$ and $U(1)$ noise :
\begin{equation}
  ( {\bf q}, {\rm noise})=\{\,
  ( {\bf 0}, \xi  ) ,
  ( {\bf 0}, \eta ) ,
  ( {\bf p}, \xi  ) ,
  (-{\bf p}, \xi  )\,
  \}
\ ,
\end{equation}
and $W$-type propagators (\ref{eq:QP_W})
for combinations of ${\bf k}$, $t_a$ and ${\bf q}$ :
\begin{equation}
   (  {\bf k}, t_a |  {\bf q} )=\{\,
   (  {\bf p}, t_S |  {\bf 0} ) ,
   ( -{\bf p}, t_S |  {\bf 0} ) ,
   (  {\bf 0}, t_S |  {\bf p} ) ,
   (  {\bf 0}, t_S | -{\bf p} ) ,
   (  {\bf 0}, t_1 |  {\bf 0} )\,
   \}
\ ,
\end{equation}
using the same $U(1)$ noise $\xi$ in common.
The $\overline{Q}$-type (\ref{eq:QP_Q_SM})
and the $\overline{W}$-type propagator (\ref{eq:QP_W_SM})
for $({\bf k}, t_a)=({\bf 0},t_1)$
are solved for the set ${\bf z}\in {\Gamma}$.
%
%-------------------------------------------------------------------------------
% @ Parameters
%
\subsection{ Simulation parameters }
Calculations in the present work
employ $N_f=2+1$ full QCD configurations
previously generated by PACS-CS
using a renormalization group improved gauge action
and a non-perturbatively
$O(a)$-improved Wilson fermion action at $\beta=1.9$
on $32^3\times 64$ lattice
($a=0.091\,{\rm fm}$, $La=2.9\,{\rm fm}$)~\cite{conf:PACS-CS}.
We choose two subsets of the PACS-CS configurations.
One of them corresponds to the hopping parameters
$\kappa_{ud}=0.13754$ for the degenerate up and down quarks and
$\kappa_{s }=0.13640$ for the strange quark,
for which the pion mass takes
$m_\pi=410\,{\rm MeV}$ ($m_\pi/m_\rho=0.46$).
The total number of configurations
analyzed every $10$ trajectories is $440$.
We estimate the statistical errors by the jackknife method
with bins of $400$ trajectories.
The other set is at
$\kappa_{ud}=0.13770$ and $\kappa_{s}=0.13640$,
corresponding to $m_\pi=300\,{\rm MeV}$ ($m_\pi/m_\rho=0.35$).
The total number of configurations of this set is $400$
and the measurements are done every $20$ trajectories.
The statistical errors
are estimated by the jackknife method
with bins of $800$ trajectories.

The periodic boundary conditions are imposed
for both spatial and temporal directions
in configuration generations.
We impose
the Dirichlet boundary condition
for the temporal direction
%%@M5--
at $t=0$ and $t=T$
in calculations of the quark propagators,
to avoid the unwanted thermal contributions
produced by propagating two pions
in opposite directions in a time.
For both quark masses,
we set the source operators
$\overline{(\pi\pi)}(t_s)$ in (\ref{eq:pp_op_source}) and
$\overline{\rho}(t_s)$ in (\ref{eq:rho_op_source})
at $t_s=12$ to avoid effects from the temporal boundary,
and the zero momentum pion in the sink operator
$(\pi\pi)(t)$ in (\ref{eq:pp_op_sink}) at $t_1=42$.

In order to see effects from the Dirichlet boundary,
we calculate the time correlation functions
of the single meson channels setting the source at $t_s=12$
with the Dirichlet boundary condition.
We compare these with those obtained
with the periodic boundary condition
in the previous work in Ref.~\cite{conf:PACS-CS},
and find no difference beyond statistical fluctuations
between them.
This shows that $t_s=12$ is sufficiently large
to avoid effects from the boundary
for the single meson channels.
We assume from this that $t_s=12$
and $t_1=42$ ($22$ away from the boundary)
are safe distances
to avoid effects from the boundary
also for the time correlation functions of the two-pion state.
We find that the effective mass
of the time correlation function
for the pion with the zero momentum
reaches plateau after $t-t_s = 12$ for the both quark masses.
We can expect from this that
the eigenvalues $\lambda_n(t)$ ($n=1,2$)
of the matrix $M(t)$ in (\ref{eq:def_M}) have
a single exponential behaviors
in a time range $t \le t_1 - 12$ ($=30$).

In a pilot study of the present work at $m_\pi=410\,{\rm MeV}$,
calculations of the phase shift
were carried out only for three representations,
${\bf T}_{1}^{-}$, ${\bf E}^{-}$ and ${\bf A}_{2}^{-}$,
setting the number of the random noise $N_R=10$ in (\ref{eq:xi_prop}).
The results on this pilot study have been reported
in Ref.~\cite{rhd:SCPH:PACS-CS}.
We found from this study that
errors from a finiteness of the number of the random noise
are small enough compared with the statistical error
for three representations
even if we take $N_R=2$.
%%@T6-- Considering that this observation applies also
Assuming that this observation applies also
for both quark masses and all representations
in the present work,
we set $N_R=2$ for all calculations
after this pilot study.

In order to reduce the statistical error,
we carry out additional measurements
shifting the time slice of the source operators $t_s$,
the zero momentum pion in the sink operator $t_1$
and the Dirichlet boundary condition
by a time shift $\Delta t$ simultaneously,
and average over these measurements.
For $m_\pi=410\,{\rm MeV}$,
the measurement of the pilot study
for the three representations
was done without shifting the time slices.
We carry out additional measurements for all representations
with the time shift $\Delta t =T/2$ and $T/4$.
We average over these two measurements
for the ${\bf B}_{1}^{-}$ representation and
all three measurements
for three representations,
${\bf T}_{1}^{-}$, ${\bf E}^{-}$ and ${\bf A}_{2}^{-}$.
For $m_\pi=300\,{\rm MeV}$,
the measurements for all representations
are carried out
with the time shift $\Delta t=0$ and $T/2$,
and are averaged.
%
%===============================================================================
%
\section{ Results }
\label{Sec:Results}
%
%-------------------------------------------------------------------------------
% @ Time correlation function
%
\subsection{ Time correlation function }
\begin{figure}[h]
\includegraphics[width=11.0cm]{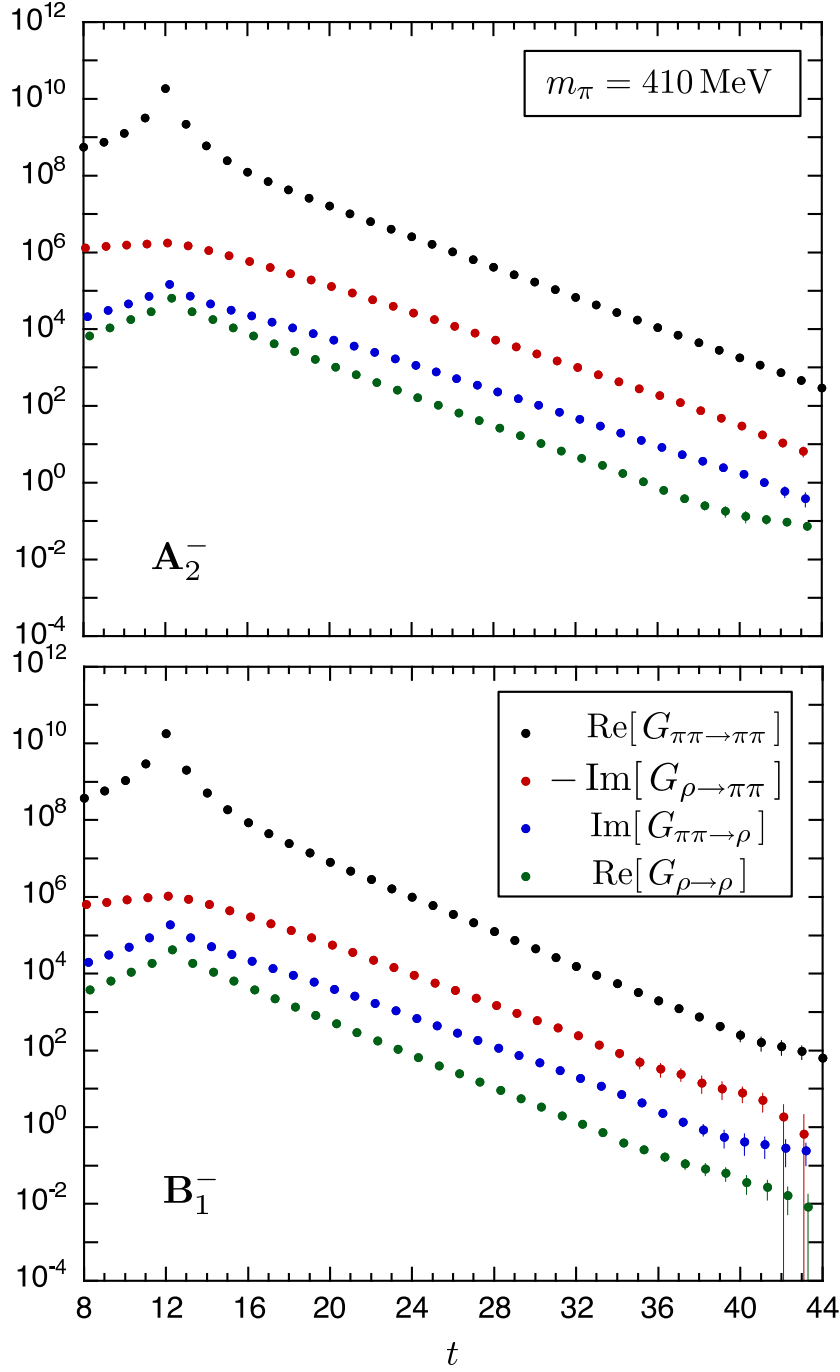}
\caption{
Four components of the matrix of the time correlation function
$G(t)$ at $m_\pi=410\,{\rm MeV}$
for the ${\bf A}_2^{-}$ (upper panel) and
for the ${\bf B}_1^{-}$ representation (lower panel).
Same symbols for the components are used in both panels.
The source operators
$\overline{(\pi\pi)}(t_s)$ and
$\overline{\rho}(t_s)$ are located at $t_s=12$.
The pion with the zero momentum in the sink operator $(\pi\pi)(t)$
in (\ref{eq:pp_op_sink}) is set at $t_1=42$.
}
\label{fig:G_410}
%------------
\newpage
%------------
\end{figure}

In Fig.~\ref{fig:G_410} we show
the real part of the diagonal components
($\pi\pi\to\pi\pi$ and $\rho\to\rho$ )
and imaginary part of the off-diagonal components
($\rho\to\pi\pi$ and $\pi\pi\to\rho$)
of the matrix of the time correlation function $G(t)$ in (\ref{eq:def_G})
for the ${\bf A}_{2}^{-}$ and the ${\bf B}_{1}^{-}$ representation
at $m_\pi=410\,{\rm MeV}$.
We note that
the diagonal components are real and
the off-diagonal components are pure imaginary
by $P$ and $CP$ symmetry.
Choosing $t_R=23$ as the reference time of the variational method
for the matrix $M(t)$ in (\ref{eq:def_M}),
we obtain the two eigenvalues $\lambda_1(t)$ and $\lambda_2(t)$
of the matrix,
which corresponds to the correlation function for the ground
and the first excited state respectively, for each representation.

\begin{figure}[h]
\includegraphics[width=11.0cm]{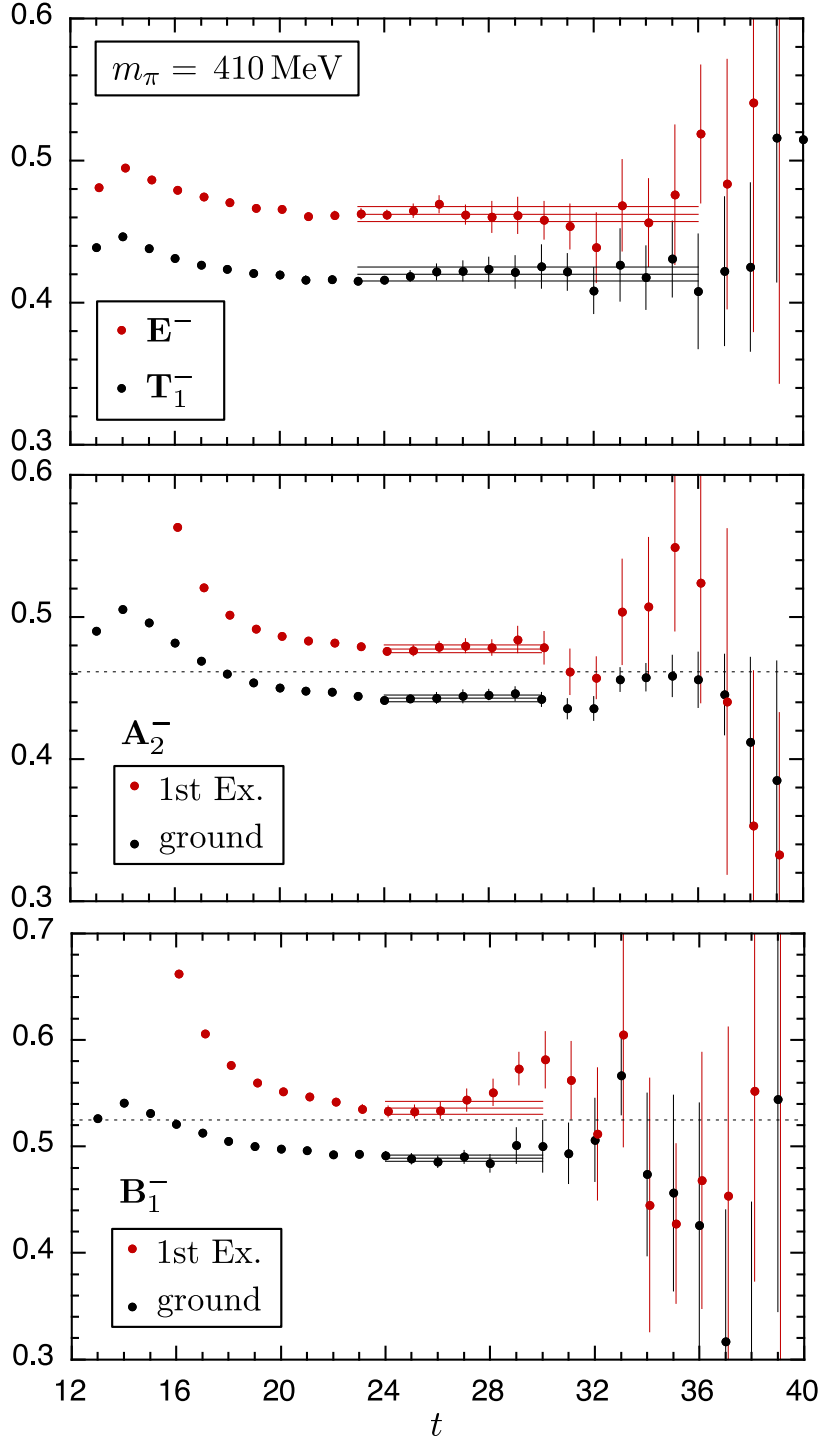}
\caption{
Effective masses of
the ground state for the ${\bf T}_1^-$ and
the ${\bf E}^-$ representation,
and the ground and first excited states
for the ${\bf A}_2^-$ and ${\bf B}_1^-$ representation
at $m_\pi=410\,{\rm MeV}$.
The source operators
$\overline{(\pi\pi)}(t_s)$ and
$\overline{\rho}(t_s)$ are located at $t_s=12$.
For the ${\bf A}_2^-$ and ${\bf B}_1^-$ representation,
we set the pion with the zero momentum
in the sink operator $(\pi\pi)(t)$
in (\ref{eq:pp_op_sink}) at $t_1=42$
and the reference time of the variational method
at $t_R=23$.
The results of the fitting
with one standard deviation error band
are expressed by solid lines.
The dotted lines for the the ${\bf A}_2^-$
and ${\bf B}_1^-$ representation
indicate the energies of the two free pions.
}
\label{fig:LM_410}
%------------
\newpage
%------------
\end{figure}

The effective masses of the time correlation functions
for six states considered in the present work
at $m_\pi=410\,{\rm MeV}$
are plotted in Fig.~\ref{fig:LM_410}.
We can find plateaus in the time region $t \ge 23$.
The results of the energies $E$
extracted
%%@T4-- by a single exponential fitting
by a single exponential fit
for these time correlation functions
are tabulated in Table~\ref{table:results_410},
together with adopted fitting ranges.
We choose smaller value for the maximum time of the fitting range
for the ${\bf A}_2^-$ and the ${\bf B}_1^-$ representation
than those for the others
to avoid contamination from higher energy states
produced by the zero momentum pion at $t_1=42$
in the sink operator $(\pi\pi)(t)$ in (\ref{eq:pp_op_sink}).
In Fig.~\ref{fig:LM_410}
the results of the fitting
with one standard deviation error band
are also expressed by solid lines.
The dotted line for the ${\bf A}_2^-$
and ${\bf B}_1^-$ representation in the figure
indicates the energy of the two free pions
for each representation.

\begin{figure}[h]
\includegraphics[width=11.0cm]{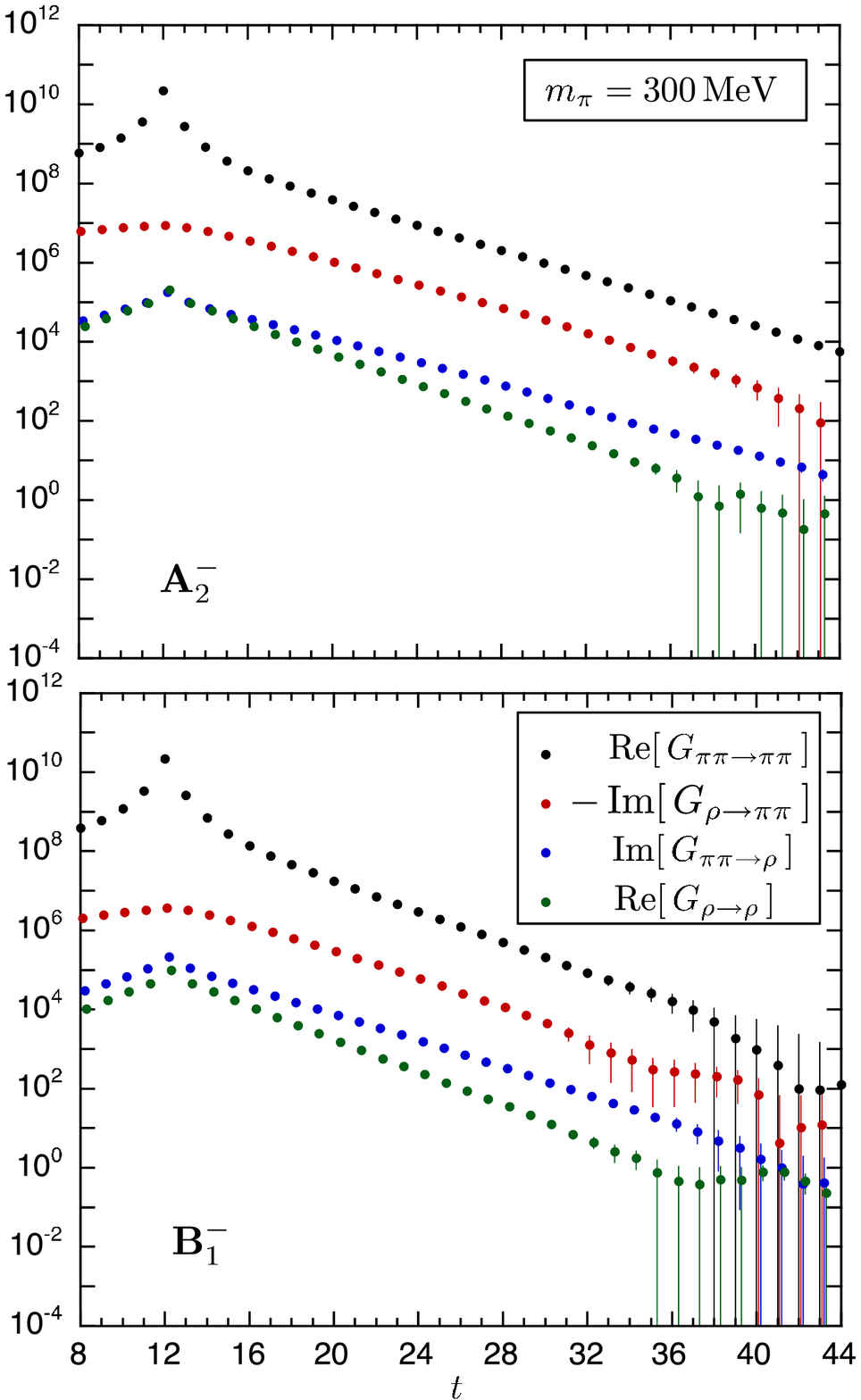}
\caption{
Same as Fig.~\ref{fig:G_410} for $m_\pi=300\,{\rm MeV}$.
}
\label{fig:G_300}
%------------
\newpage
%------------
\end{figure}

\begin{figure}[h]
\includegraphics[width=11.0cm]{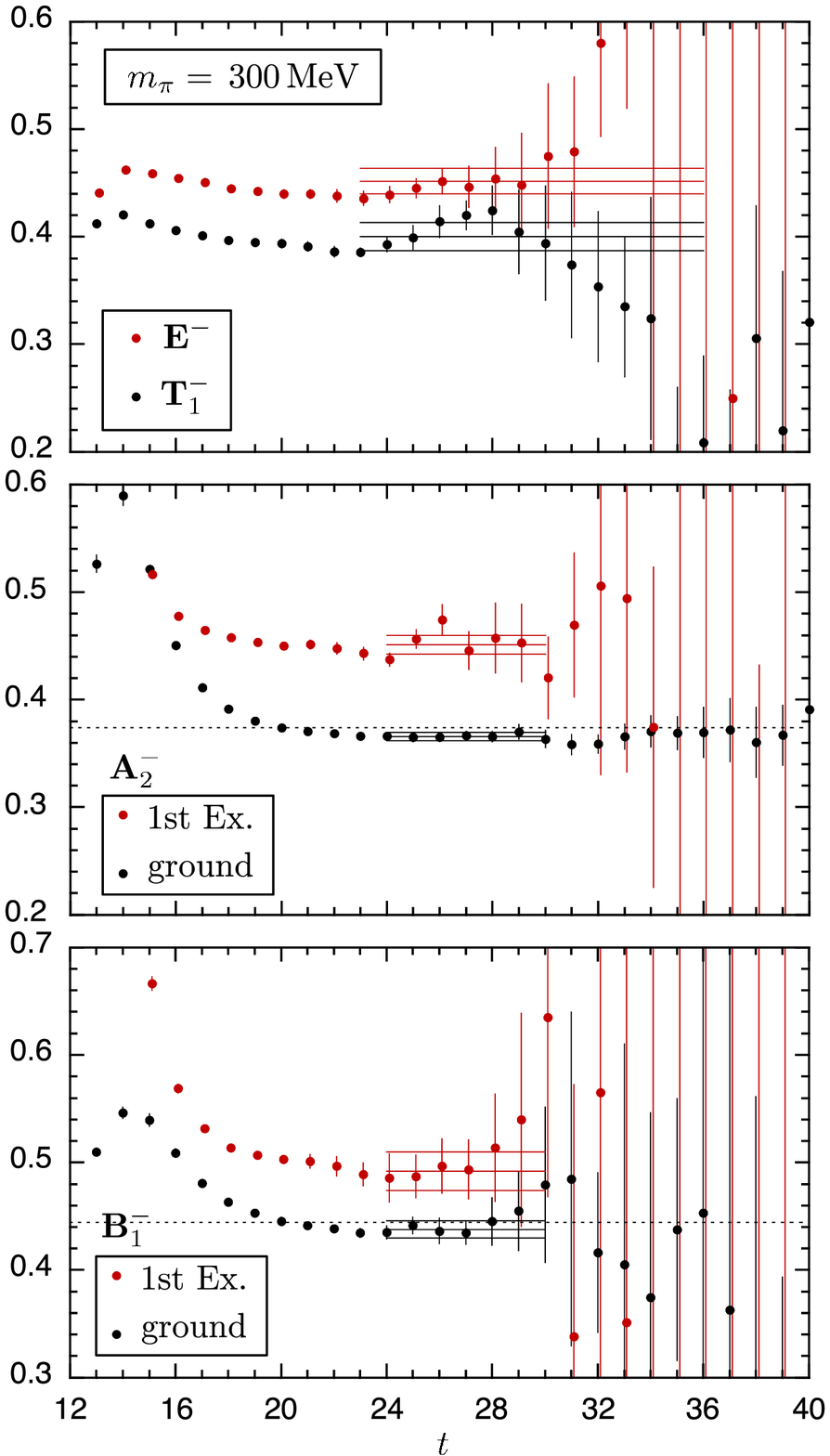}
\caption{
Same as Fig.~\ref{fig:LM_410} for $m_\pi=300\,{\rm MeV}$.
}
\label{fig:LM_300}
%------------
\newpage
%------------
\end{figure}

The components of the matrix of the time correlation function $G(t)$
at $m_\pi=300\,{\rm MeV}$
are plotted in Fig.~\ref{fig:G_300} and
the effective masses in Fig.~\ref{fig:LM_300},
where $t_R=23$ is also chosen as the reference time.
The statistics is less than that at $m_\pi=410\,{\rm MeV}$,
but we also see plateaus in the effective masses for $t \ge 23$.
The results of the energies $E$
extracted by a single exponential fit
are tabulated in Table~\ref{table:results_300}.

In the previous work by CP-PACS,
carried out at the lattice spacing $a=0.21\,{\rm fm}$,
they found a large violation of the continuum dispersion relation
for the single pion state
due to the discretization error on their gauge configurations.
The discretization error also affects calculations of
the invariant mass $\sqrt{s}$
and the scattering momentum $k$ for the two-pion system
since they are evaluated from the energy $E$.
The continuum relation is given by
$\sqrt{s}=\sqrt{E^2 - |{\bf P}|^2}=2\sqrt{k^2+m_\pi^2}$, while
there are several alternatives on the lattice, {\it e.g., }
\begin{eqnarray}
&&  {\rm cosh}( \sqrt{s} ) = {\rm cosh}(E) - 2 \sum_{j=1}^{3} \sin^2(P_j/2)
\ ,
\label{eq:Lat_D}
\\
&&
  2 \cdot \sin^2(k/2) = {\rm cosh }(\sqrt{s}/2) - {\rm cosh}(m_\pi)
\ .
\label{eq:Lat_D_k2}
\end{eqnarray}
The two-pion scattering momentum cannot be uniquely defined
due to the breaking of the translational and rotational symmetries
in the finite lattice spacing as mentioned in Ref.~\cite{Lfm:RG}.
The momentum $k$ given by (\ref{eq:Lat_D_k2}) is just one of the choices
of the momentum, thus
the discretization error cannot be fully avoided
by using (\ref{eq:Lat_D}) and (\ref{eq:Lat_D_k2}).
In the work by CP-PACS,
they regarded the difference of the final results
for the choice of the relations
as the systematic error from the discretization error.

We also monitor
the validity of the continuum dispersion relation
for the single pion state and find that the violation is negligible
in the present work,
one reason being that our lattice spacing
$a=0.091\,{\rm fm}$ is much smaller than that for the CP-PACS case.
We compare the energy $E$ extracted from the
the time correlation function with that given
by the dispersion relation $E_{\rm eff}=\sqrt{ |{\bf p}|^2 + m_\pi^2}$
from the mass $m_\pi$ and the momentum ${\bf p}$.
The results for $m_\pi=410\,{\rm MeV}$ are
$E/E_{\rm eff} = 0.9988(15)$ for ${\bf p}=(2\pi/L) (0,0,1)$
and             $0.9988(63)$ for ${\bf p}=(2\pi/L) (1,1,0)$.
Those for $m_\pi=300\,{\rm MeV}$ are
$0.9924(81)$ and $0.984(16)$.
Therefore, the violation of the dispersion relation
for the single pion state is not seen
on our gauge configurations.
We also evaluate $\sqrt{s}$ for the two-pion state by (\ref{eq:Lat_D}),
but we see no difference over the statistics from
that given by the relation in the continuum.
From this we calculate
$\sqrt{s}$ and $k$
from $E$ by the relation in the continuum,
avoiding ambiguities possibly caused by the choice of the relations.
The results given by this way are tabulated in
Table~\ref{table:results_410} and
Table~\ref{table:results_300}.
%
%-------------------------------------------------------------------------------
% @ Scattering phase shift and resonance parameters }
%
\subsection{ Scattering phase shift and resonance parameters }
\begin{figure}[h]
\includegraphics[width=11.0cm]{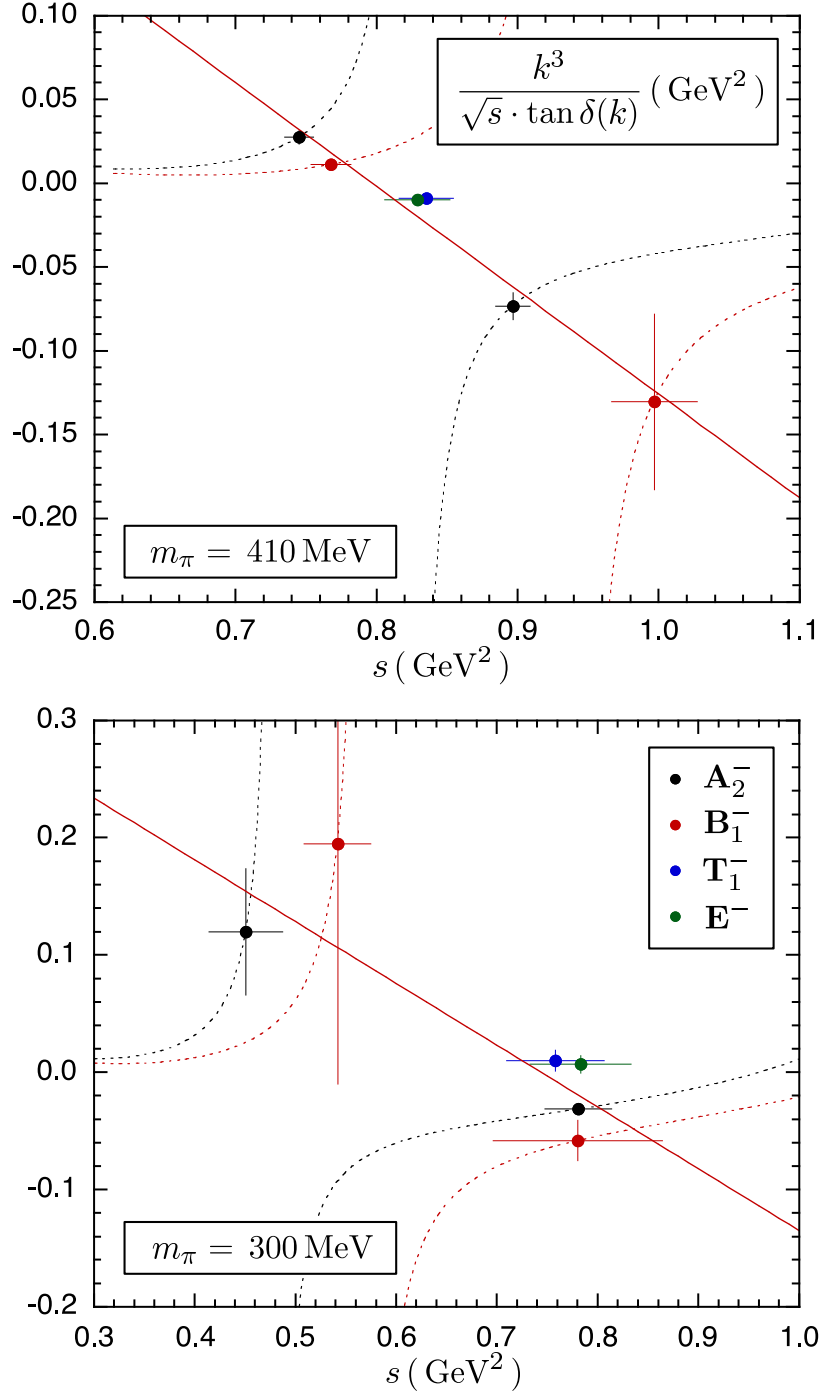}
\caption{
$k^3/(\sqrt{s}\cdot\tan\delta(k))$
as a function of square of the invariant mass ${s}$
at  $m_\pi=410\,{\rm MeV}$ (upper panel)
and $m_\pi=300\,{\rm MeV}$ (lower panel).
Same symbols for four representations are used in both panels.
Dotted lines are the finite size formulas for
the ${\bf A}_2^{-}$ and the ${\bf B}_1^{-}$ representation.
A solid line for each quark mass is a fitting line
by (\ref{eq:k2_AMP_ERF}).
}
\label{fig:k2_AMP_SS}
%------------
\newpage
%------------
\end{figure}

The scattering phase shift $\delta(k)$ obtained by substituting
the scattering momentum $k$ and the total momentum ${\bf P}$
into the finite size formulas in (\ref{eq:FSF_our_rep}) are
presented in the lower part of
Table~\ref{table:results_410} for $m_\pi=410\,{\rm MeV}$ and
Table~\ref{table:results_300} for $m_\pi=300\,{\rm MeV}$.
We use the lattice spacing determined
from $m_\Omega$ in Ref.~\cite{conf:PACS-CS},
$a=0.0907(13)\,{\rm fm}$ ($1/a=2.176(31)\,{\rm GeV}$),
to get the values in the physical unit,
where the error of the lattice spacing is not included.
In Fig.~\ref{fig:k2_AMP_SS}
the results of $k^3/(\sqrt{s}\cdot\tan\delta(k))$ are plotted
as a function of square of the invariant mass $s$ for
$m_\pi=410\,{\rm MeV}$ (upper panel) and
$m_\pi=300\,{\rm MeV}$ (lower panel).
The finite size formulas for the ${\bf A}_2^-$
and the ${\bf B}_1^-$ representation
are plotted by dotted lines.
Divergent point
on these lines corresponds to the square of the invariant mass
of the two free pions.
In the figure the error bars
of $s$ and $k^3/(\sqrt{s}\cdot\tan\delta(k))$
are plotted regarding them  as independent.
But these are fully correlated by the finite size formula,
so the true error lies
along the dotted line corresponding to the formula.
Thus, the error bars in the figure
indicate projections of the true error bar
on the finite size formula
to the vertical and the horizontal axis.

In order to extract the resonance parameters from
the results of the scattering phase shift,
we try to parametrize the resonant behavior
of the $P$-wave phase shift
in terms of the effective $\rho\to\pi\pi$ coupling constant
$g_{\rho\pi\pi}$ as
\begin{equation}
  \frac{k^3}{\tan\delta(k)} / \sqrt{s}
  = \frac{6\pi}{g_{\rho\pi\pi}^2}
    \,( m_\rho^2 - s )
\ ,
\label{eq:k2_AMP_ERF}
\end{equation}
where $m_\rho$ is the resonance mass and $g_{\rho\pi\pi}$ is defined
though the effective $\rho\to\pi\pi$ Lagrangian as
\begin{equation}
  L_{\rm eff} = g_{\rho\pi\pi}
    \sum_{\mu abc} \epsilon_{abc}
    ( k_1 - k_2 )_\mu  \ \rho_\mu^a(p) \pi^b(k_1) \pi^c(k_2)
\ .
\label{eq:def_gRhPiPi}
\end{equation}
This parametrization has been widely used
in the previous works of the $\rho$ meson decay.
The $\rho$ meson decay width at the physical quark mass
is related to the coupling constant by
\begin{eqnarray}
  \Gamma_\rho
   = \frac{g_{\rho\pi\pi}^2}{6\pi}
     \frac{ \left( k^{\rm ph}\right) ^3}{m^{\rm ph}_\rho}
   = 4.237\,{\rm MeV} \times g_{\rho\pi\pi}^2
\ ,
\label{eq:GR_gRhPiPi}
\end{eqnarray}
where $m^{\rm ph}_{\rho}=775.5\,{\rm MeV}$
is the actual $\rho$ meson mass
and $\left(k^{\rm ph}\right)^2
= \left( m^{\rm ph}_\rho \right)^2 / 4 - \left( m^{\rm ph}_{\pi} \right)^2$
($m^{\rm ph}_{\pi}=135\,{\rm MeV}$).

By chi-square fitting of the scattering phase shifts
with the fit function (\ref{eq:k2_AMP_ERF}),
we obtain,
\begin{eqnarray}
 g_{\rho\pi\pi} &=& 5.52   \pm 0.40    \ , \cr
 a m_\rho       &=& 0.4103 \pm 0.0026  \ , \cr
   m_\rho       &=& 892.8  \pm 5.5  \pm 13    \,{\rm MeV}
\ ,
\label{eq:result_410}
\end{eqnarray}
for $m_\pi=410\,{\rm MeV}$,
where the first error of $m_\rho$
is the statistical and
the second is the systematic uncertainty
for the determination of the lattice spacing.
In the fitting,
we define the chi-square for each data point
by squaring the ratio
of the distance from the data point to the fitting line
(\ref{eq:k2_AMP_ERF}) along the finite size formula
and the true statistical error calculated along the finite size formula.
The errors of the resonance parameters $g_{\rho\pi\pi}$ and $m_\rho$
are estimated by the jackknife method as for the other values.
In the upper panel of Fig.~\ref{fig:k2_AMP_SS}
we draw a fitting line by a solid line.
We can find that
the fitting with the function (\ref{eq:k2_AMP_ERF})
goes well in the large energy region
at $m_\pi=410\,{\rm MeV}$.

For $m_\pi=300\,{\rm MeV}$
the statistics of our data is not enough to discuss
a quality of the fitting with the fit function (\ref{eq:k2_AMP_ERF})
as shown in Fig.~\ref{fig:k2_AMP_SS}.
Improving the statistic by using some efficient smearing techniques
for the two-pion operator may be necessary for an investigation
of a reliability of (\ref{eq:k2_AMP_ERF}).
We must leave this issue to studies in the future.
Here
we carry out the chi-square fitting
as done at $m_\pi=410\,{\rm MeV}$,
assuming that the function (\ref{eq:k2_AMP_ERF})
also works well in our energy region at $m_\pi=300\,{\rm MeV}$.
The results of the fitting are given by
\begin{eqnarray}
 g_{\rho\pi\pi} &=& 5.98  \pm 0.56   \ , \cr
 a m_\rho       &=& 0.396 \pm 0.010  \ , \cr
   m_\rho       &=& 863   \pm 23 \pm 12    \,{\rm MeV}
\ ,
\label{eq:result_300}
\end{eqnarray}
where the second error of $m_\rho$
is the systematic uncertainty
for the determination of the lattice spacing.
We draw a fitting line
by a solid line in the lower panel of Fig.~\ref{fig:k2_AMP_SS}.

From (\ref{eq:result_410}) and (\ref{eq:result_300})
we find that the $g_{\rho\pi\pi}$ at the two quark masses
are consistent within the statistical error and also
with the experiment $g_{\rho\pi\pi}=5.874\pm 0.014$
given from the experimental result of the decay width
$\Gamma_\rho = 146.2 \pm0.7\,{\rm MeV}$~\cite{PDG:2010}
by (\ref{eq:GR_gRhPiPi}).
This suggests a weak quark mass dependence
of the coupling constant.
But our calculations are carried out only at the two quark masses
and a reliability of (\ref{eq:k2_AMP_ERF})
is assumed in the analysis at $m_\pi=300\,{\rm MeV}$,
so high statistical calculations
at more quark masses are necessary
to obtain a definite conclusion for the quark mass dependence.
We also leave this issue to studies in the future.
%
%-------------------------------------------------------------------------------
% @ Comparison with other works
%
\subsection{ Comparison with other works }
\begin{figure}[h]
\includegraphics[width=11.0cm]{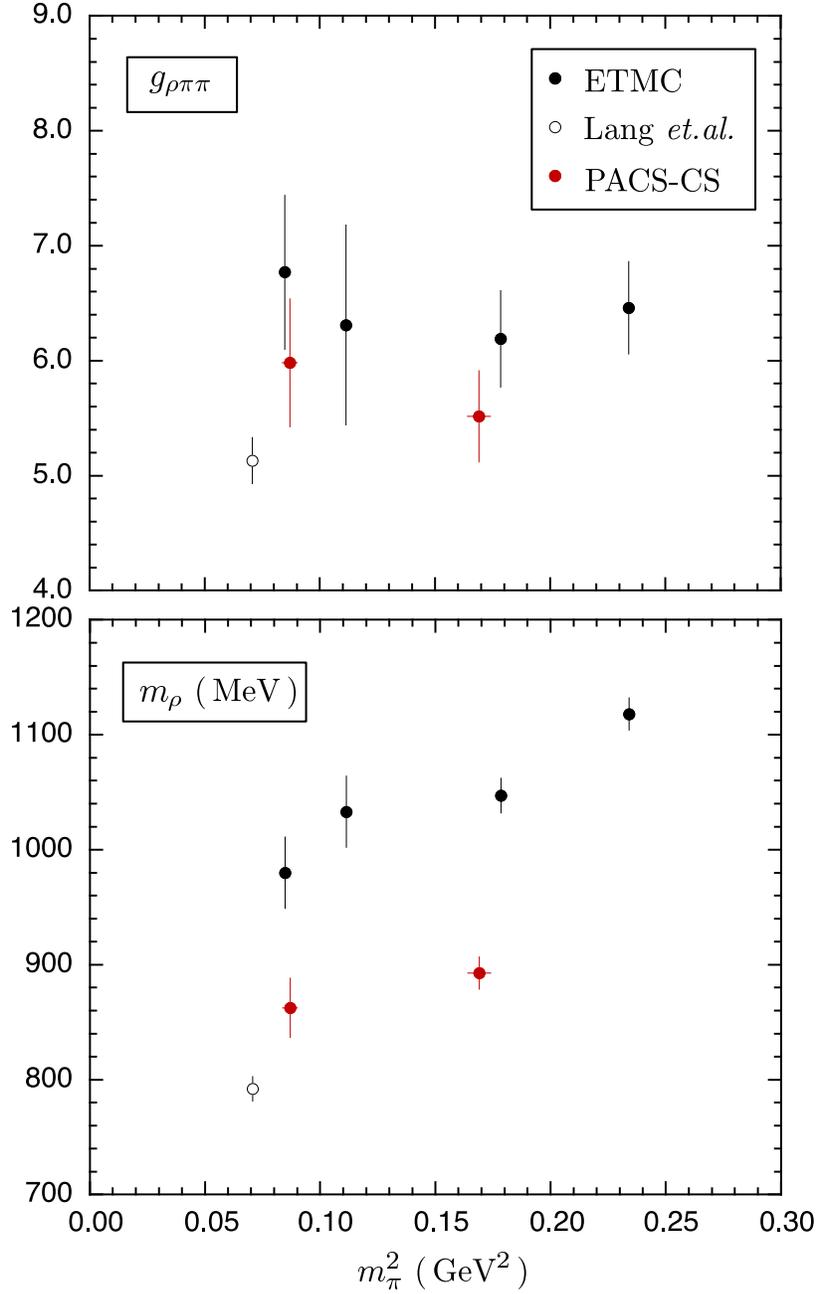}
\caption{
Comparison of our results (PACS-CS) obtained in $2+1$ flavor QCD
with those by ETMC
and Lang {\it et al.} in $2$ flavor QCD.
Upper panel shows
the effective coupling constant $g_{\rho\pi\pi}$ and
lower is the resonance mass $m_\rho$.
The systematic uncertainty
for the determination of the lattice spacing
is added to the statistical error in quadrature.
}
\label{fig:comp}
%------------
\newpage
%------------
\end{figure}

In Fig.~\ref{fig:comp}
we compare
our results (PACS-CS) obtained  in $2+1$ flavor QCD
with those by ETMC~\cite{rhd:SCPH:ETMC_1,rhd:SCPH:ETMC_2}
and Lang {\it et al.}~\cite{rhd:SCPH:LANG} in $2$ flavor QCD.
The upper panel shows
the effective coupling constant $g_{\rho\pi\pi}$
and the lower panel displays the resonance mass $m_\rho$
as a function of $m_\pi^2$.
Here the systematic uncertainty
for the determination of the lattice spacing
is added to the statistical error in quadrature.
A good agreement between our result and ETMC is
observed for $g_{\rho\pi\pi}$.
The result for the coupling constant
by Lang {\it et al.} takes a slightly  smaller value,
but it is almost consistent with other works.

\begin{figure}[h]
\includegraphics[width=11.0cm]{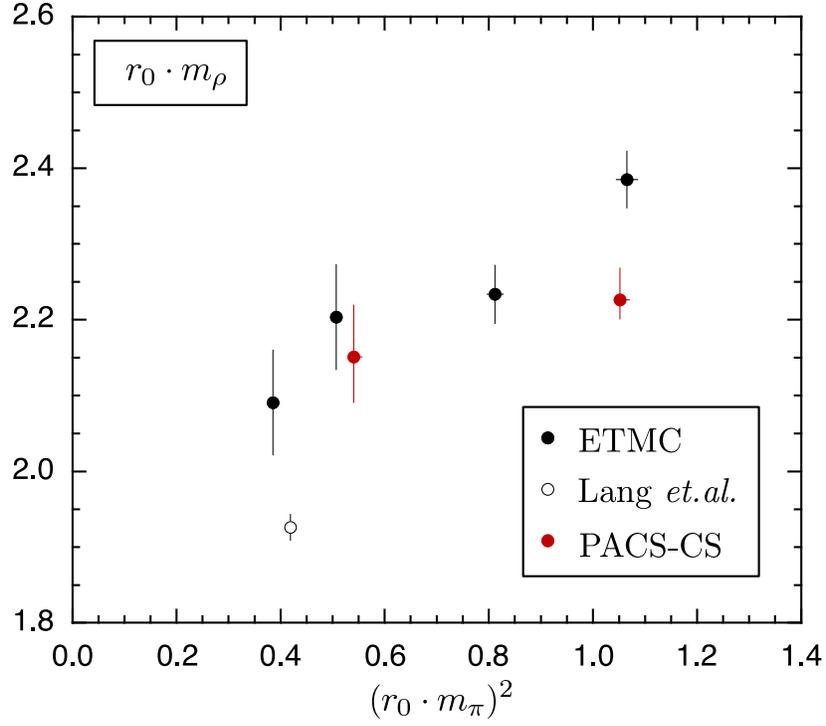}
\caption{
Comparison of our results (PACS-CS) with those
by ETMC and Lang {\it et al.}
for dimensionless value $r_0 m_\rho$ as a function of $(r_0 m_\pi)^2$
with the Sommer scale $r_0$.
The error of $r_0$
is added to the statistical error in quadrature.
}
\label{fig:comp_mRr0}
%------------
\newpage
%------------
\end{figure}

We see, however, large discrepancy for the resonance mass $m_\rho$
in the lower panel of Fig.~\ref{fig:comp}.
One of possible reason for this discrepancy is the systematic error
from the determination of the lattice spacing
which is used to obtain $m_\rho$ and $m_\pi$ in the physical unit.
In the present work
the lattice spacing
$a=0.0907(13)\,{\rm fm}$ determined from $m_\Omega$
in Ref.~\cite{conf:PACS-CS} is used as explained before.
ETMC used $a=0.079(2)(\pm2)\,{\rm fm}$ given
from the pion decay constant $f_\pi$ in Ref.~\cite{ETMC-conf}.
In the work by Lang {\it et al.},
the authors determined it to be $a=0.1239(13)\,{\rm fm}$
from the Sommer scale $r_0=0.48\,{\rm fm}$ as input.
In order to avoid a spurious systematic error
from the determination of lattice spacing,
it is appropriate to compare our results
with other works in terms of dimensionless quantities.
In Fig.~\ref{fig:comp_mRr0}
we plot $r_0 m_\rho$ as a function of $(r_0 m_\pi)^2$
with the Sommer scale $r_0$.
The value of $r_0$
for the PACS-CS configurations
has been reported as
$r_0 / a = 5.427(51)(+81)(-2)$~\cite{conf:PACS-CS}
and that for ETMC
as $r_0/a=5.32(5)$~\cite{ETMC-conf}.
In the figure
the statistical error
and the systematic uncertainty
for the determination of $r_0$
are added in quadrature.
We see that the discrepancy
between ours and ETMC tends to be smaller,
but it still remains for the large quark mass.
The result by Lang {\it et al.}
takes a smaller value than those of the two works.
The finite size effect can be considered
as a possible reason of their small value of $m_\rho$
as commented by themselves  in their paper.
Their lattice extent $La=1.98\,{\rm fm}$
may not be large enough for
their quark mass $m_\pi=266\,{\rm MeV}$.
The three groups worked at a single lattice spacing,
therefore the another possible reason of
the discrepancy is the discretization error
due to the finite lattice spacing.
We can also consider several other reasons,
the dynamical strange quark effect,
the isospin breaking effect,
the reliability of the parametrization of the scattering phase shift
by (\ref{eq:k2_AMP_ERF}) and so on,
but a definite conclusion can not be given here.
A precise determination of the resonance mass $m_\rho$
by the calculation near or on the physical point closer
to the continuum limit
is an important work reserved for the future.
%
%===============================================================================
% @ Conclusions
%
\section{ Conclusions }
\label{Sec:Conclusions}
We have reported on a calculation of the $P$-wave scattering phase shift
for the isospin $I=1$ two-pion system
and estimations of the resonance parameters of the $\rho$ meson
from the $N_f=2+1$ full QCD configurations
with a large lattice volume.
The calculations are carried out at two quark masses,
which correspond to $m_\pi=410\,{\rm MeV}$ and $300\,{\rm MeV}$.

In order to extract the resonance parameters from
the scattering phase shift,
we parametrize the resonant behavior
of the $P$-wave phase shift
in terms of the effective coupling constant $g_{\rho\pi\pi}$
and the resonance mass $m_\rho$.
We find that this parametrization works well
in the large energy region
for our data at $m_\pi=410\,{\rm MeV}$
and obtain $g_{\rho\pi\pi} = 5.52 \pm 0.40$.

For $m_\pi=300\,{\rm MeV}$
the statistics of our data is not enough
to discuss the reliability of the parametrization.
We leave an investigation on this point
to the studies in the future.
We carry out the fitting
assuming that this parametrization also works
in our energy region at $m_\pi=300\,{\rm MeV}$.
Our result is $g_{\rho\pi\pi} = 5.98 \pm 0.56$,
which agrees with the coupling constant at $m_\pi=410\,{\rm MeV}$
and the experiment $g_{\rho\pi\pi}=5.874\pm 0.014$
within the statistical error.
This suggests a weak quark mass dependence
of the coupling constant.
The studies at more quark masses
are necessary to obtain a definite conclusion
for the quark mass dependence, however.

We find a discrepancy for the resonance mass $m_\rho$
among three lattice studies.
Although a part of the discrepancy seems to be explained
by different choices of the scale setting,
other sources such as
the discretization error due to the finite lattice spacing,
the dynamical strange quark effect,
the isospin breaking effect and
the reliability of the parametrization of the scattering phase shift
may be needed to resolve this discrepancy.
Calculations near or on the physical point closer to the continuum limit
are necessary for a precise determination
of the resonance mass from lattice QCD.
We leave this issue to studies in the future.
%
%===============================================================================
% @ Acknowledgements
%
\section*{Acknowledgments}
This work is supported in part by Grants-in-Aid
of the Ministry of Education
(Nos.
%
% aoki      :
20340047, 20105001, 20105003,
%
% ishizuka  :
20540248, 23340054,
%
% kanaya    :
21340049,
%
% kuramashi :
22244018, 20105002,
%
% namekawa  :
22105501, 22740138,
%
% okawa :
23540310,
%
% taniguchi :
22540265, 23105701,
%
% ukawa    :
18104005,
%
% yamazaki  :
21105501, 23105708,
%
% yoshie :
20105005 ).
The numerical calculations have been carried out
on PACS-CS at Center for Computational Sciences, University of Tsukuba.
%
%==========================================================================
%

%
%======================================================================
%
\newpage
\appendix
%
%======================================================================
% @ Figure
%
%
%=====================================================================
%
%
%=====================================================================
%
%
%=====================================================================
%
%
%=====================================================================
%
%
%=====================================================================
%
%
%=====================================================================
%
%
%=====================================================================
%
%
%=====================================================================
%
%
%=====================================================================
% @ Table
%
\begin{table}[t]
\caption{
The ground and the first excited states
with the isospin $(I,I_z)=(1,0)$
for the irreducible representations
considered in the present work,
ignoring the hadron interactions.
${\bf P}$ is the total momentum,
${\rm g}$ is the rotational group in each momentum frame on the lattice
and $\Gamma$ is the irreducible representation of the rotational group.
The vectors in parentheses after
$(\pi\pi)$ and $\rho$ refer
to the momenta of the two pions and the $\rho$ meson
in unit of $2\pi/L$.
We use a notation
$  (\pi\pi)({\bf p}_1)({\bf p}_2)
  =    \pi^{+} ({\bf p}_1) \pi^{-}({\bf p}_2)
     - \pi^{-} ({\bf p}_1) \pi^{+}({\bf p}_2)
$ for the two-pion states.
An index $j$ for the ${\bf T}_1^{-}$ representation
takes $j=1,2,3$ and
$k$ for the ${\bf E}^{-}$ takes $k=1,2$.
}
\label{table:calc_rep}
\begin{ruledtabular}
\begin{tabular}{llll ll}
frame & ${\bf P} L/(2\pi)$
& g
& $\Gamma$ &
\\
\hline
%---------------------------------------------------
CMF  & $(0,0,0)$  & $O_h$ & ${\bf T}_{1}^{-}$
& $\rho_{j}({\bf 0})$
& $(\pi\pi)({\bf e}_j)(-{\bf e}_j)$
\\
%---------------------------------------------------
MF1  & $(0,0,1)$  & $D_{4h}$ & ${\bf E}^{-}$
& $\rho_{k}(0,0,1)$
& $(\pi\pi)({\bf e}_3 + {\bf e}_k)
           (          - {\bf e}_k)
  -(\pi\pi)({\bf e}_3 - {\bf e}_k)
           (            {\bf e}_k)$
\\
%--------------------------------------------------
MF1  & $(0,0,1)$  & $D_{4h}$ & ${\bf A}_{2}^{-}$
& $\rho_{3}(0,0,1)$
& $(\pi\pi)(0,0,1)(0,0,0)$
\\
%--------------------------------------------------
MF2  & $(1,1,0)$  & $D_{2h}$ & ${\bf B}_{1}^{-}$
& $(\rho_{1}+\rho_{2})(1,1,0)$
& $(\pi\pi)(1,1,0)(0,0,0)$
\\
%-----------------------------------
\end{tabular}
\end{ruledtabular}
%
%----------------
\newpage
%----------------
\end{table}
%
%=====================================================================
% @ Table
%
\begin{table}[t]
\caption{
Results at $m_\pi=410\,{\rm MeV}$.
${\bf P}$ is the total momentum
and $\Gamma$ is the irreducible representation
of the rotational group on the lattice.
$E$ is the energy extracted by fitting the time correlation function
with the fitting range in a line of ``Fit Range''.
$\sqrt{s}$ is the invariant mass and
$k$ is the scattering momentum,
which are related
by $\sqrt{s} = \sqrt{ E^2 - |{\bf P}|^2 } = 2\sqrt{k^2 + m_\pi^2}$.
$\delta(k)$ is the $P$-wave scattering phase shift
given by the finite size formulas in (\ref{eq:FSF_our_rep}).
We use the value of the lattice spacing
given in the previous work in Ref.~\cite{conf:PACS-CS},
$a=0.907(13)\,{\rm fm}$ ($1/a=2.176(31)\,{\rm GeV}$),
to obtain the values in the physical unit,
where the error of the lattice spacing is not included.
}
\label{table:results_410}
\begin{ruledtabular}
\begin{tabular}{l llllll}
$am_\pi$               & $0.18897(79)$    \\
$m_\pi\,({\rm MeV})$   & $411.2(1.7)  $    \\
\hline
%-----------------------------------
frame
&  CMF
&  MF1
&  MF1
&  %--
&  MF2
&  %--
\\
%-----------------------------------
${\bf P}L/(2\pi)$
&  $(0,0,0)$
&  $(0,0,1)$
&  $(0,0,1)$
&  %-- $(0,0,1)$
&  $(1,1,0)$
&  %-- $(1,1,0)$
\\
%-----------------------------------
$\Gamma$
&  ${\bf T}_1^-$
&  ${\bf E}^-$
&  ${\bf A}_2^-$
&  %-- ${\bf A}_2^-$
&  ${\bf B}_1^-$
&  %-- ${\bf B}_1^-$
\\
%-----------------------------------
Fit Range
& $23-36$
& $23-36$
& $24-30$
& $24-30$
& $24-30$
& $24-30$
\\
%-----------------------------------
$aE$
&  $0.4200(49)$
&  $0.4622(53)$
&  $0.4426(24)$
&  $0.4774(27)$
&  $0.4891(30)$
&  $0.5364(60)$
\\
%-----------------------------------
$a \sqrt{s}$
&  $0.4200(49)$
&  $0.4184(58)$
&  $0.3967(27)$
&  $0.4352(30)$
&  $0.4027(37)$
&  $0.4589(70)$
\\
%-----------------------------------
$(ak)^2\,\,( \times 10^{-3} )$
&  $  8.4(1.1)  $
&  $  8.1(1.2)  $
&  $  3.63(41)  $
&  $ 11.63(57)  $
&  $  4.83(80)  $
&  $ 16.9(1.5)  $
\\
%-----------------------------------
$\sqrt{s}\,({\rm MeV})$
&  $ 914(11)     $
&  $ 911(13)     $
&  $ 863.2(5.9)  $
&  $ 946.9(6.4)  $
&  $ 876.2(8.1)  $
&  $ 999(15)     $
\\
%-----------------------------------
$s \,({\rm GeV}^2)$
&  $ 0.835(19) $
&  $ 0.829(23) $
&  $ 0.745(10) $
&  $ 0.897(12) $
&  $ 0.768(14) $
&  $ 0.997(30) $
\\
%-----------------------------------
$a^2 k^3 / \tan\delta / \sqrt{s} \,\,( \times 10^{-3} )$
& $    -1.872(19) $
& $    -2.120(22) $
& $     5.79(85)  $
& $   -15.5(1.7)  $
& $     2.36(49)  $
& $   -28(11)     $
\\
%-----------------------------------
$k^3 / \tan\delta / \sqrt{s} \,\,( \times 10^{-2}\,{\rm GeV^2})$
& $   -0.8865(88)$
& $   -1.004(11) $
& $    2.74(40)  $
& $   -7.34(80)  $
& $    1.11(23)  $
& $  -13.0(5.2)  $
\\
%-----------------------------------
\end{tabular}
\end{ruledtabular}
\end{table}
%
%======================================================================
% @ Table
%
\begin{table}[t]
\caption{
Same as Table~\ref{table:results_410} for $m_\pi=300\,{\rm MeV}$.
}
\label{table:results_300}
\begin{ruledtabular}
\begin{tabular}{l llllll}
$am_\pi$               & $0.1355(15)$ \\
$m_\pi\,({\rm MeV})$   & $294.9(3.3)$  \\
\hline
%-----------------------------------
frame
&  CMF
&  MF1
&  MF1
&  %--
&  MF2
&  %--
\\
%-----------------------------------
${\bf P}L/(2\pi)$
&  $(0,0,0)$
&  $(0,0,1)$
&  $(0,0,1)$
&  %-- $(0,0,1)$
&  $(1,1,0)$
&  %-- $(1,1,0)$
\\
%-----------------------------------
$\Gamma$
&  ${\bf T}_1^-$
&  ${\bf E}^-$
&  ${\bf A}_2^-$
&  %-- ${\bf A}_2^-$
&  ${\bf B}_1^-$
&  %-- ${\bf B}_1^-$
\\
%-----------------------------------
Fitting Range
& $23-36$
& $23-36$
& $24-30$
& $24-30$
& $24-30$
& $24-30$
\\
%-----------------------------------
$aE$
&  $ 0.400(13)   $
&  $ 0.452(12)   $
&  $ 0.3658(40)  $
&  $ 0.4511(85)  $
&  $ 0.4377(80)  $
&  $ 0.492(18)   $
\\
%-----------------------------------
$a \sqrt{s}$
&  $ 0.400(13)   $
&  $ 0.407(13)   $
&  $ 0.3086(47)  $
&  $ 0.4061(95)  $
&  $ 0.338(10)   $
&  $ 0.406(22)   $
\\
%-----------------------------------
$(ak)^2\,\,( \times 10^{-3} )$
&  $   21.7(2.5) $
&  $   23.0(2.7) $
&  $    5.45(59) $
&  $   22.9(1.9) $
&  $   10.3(1.9) $
&  $   22.8(4.5) $
\\
%-----------------------------------
$\sqrt{s}\,({\rm MeV})$
&  $  871(28)  $
&  $  885(29)  $
&  $  672(10)  $
&  $  884(21)  $
&  $  736(23)  $
&  $  883(48)  $
\\
%-----------------------------------
$s\,({\rm GeV}^2)$
&  $  0.758(48)  $
&  $  0.783(50)  $
&  $  0.451(14)  $
&  $  0.781(37)  $
&  $  0.542(33)  $
&  $  0.780(84)  $
\\
%-----------------------------------
$a^2 k^3/\tan\delta /\sqrt{s} \,\,( \times 10^{-3} )$
&  $    2.1(1.9) $
&  $    1.4(1.6) $
&  $   25(11)    $
&  $   -6.60(97) $
&  $    4.1(4.3) $
&  $  -12.3(3.7) $
\\
%-----------------------------------
$k^3/\tan\delta/\sqrt{s} \,\,( \times 10^{-2}\,{\rm GeV}^2 )$
&  $    1.00(90) $
&  $    0.68(76) $
&  $   12.0(5.4) $
&  $   -3.13(46) $
&  $   19(21)    $
&  $   -5.8(1.7) $
\\
%-----------------------------------
\end{tabular}
\end{ruledtabular}
\end{table}
%
%======================================================================
%
\end{document}